\documentclass[utf8]{FrontiersinHarvard} 

\usepackage{url,hyperref,microtype,subcaption}
\usepackage[onehalfspacing]{setspace}

\usepackage{multirow}
\usepackage[dvipsnames]{xcolor}
\usepackage{soul}

\newcommand{\hexp} {{\it HEX-P}}
\newcommand{\xmm} {{\it XMM-Newton}}
\newcommand{\chandra} {{\it Chandra}}
\newcommand{\nustar} {{\it NuSTAR}}
\newcommand{\swift} {{\it Swift}}
\newcommand{\nicer} {{\it NICER}}

\newcommand{\cmsq} {cm$^{-2}$}
\newcommand{\nh} {$N_{\rm{H}}$}

\newcommand{\arcmin}{$^{\prime}$}

\newcommand{\ergs}{\mbox{\thinspace erg\thinspace s$^{-1}$}}
\newcommand{\ergcms}{\mbox{\thinspace erg\thinspace cm$^{-2}$\thinspace s$^{-1}$}}

\newcommand{\msol} {$M_{\odot}$}
\newcommand{\cow} {AT\,2018cow}

\def\keyFont{\fontsize{8}{11}\helveticabold }
\def\firstAuthorLast{Brightman {et~al.}} 
\def\Authors{
Murray Brightman\,$^{1,*}$, 
Raffaella Margutti\,$^{2,3}$, 
Ava Polzin\,$^{4}$, 
Amruta Jaodand\,$^{1}$, 
Kenta Hotokezaka\,$^{5}$, 
Jason A. J. Alford\,$^{6}$, 
Gregg Hallinan\,$^{1}$,
Elias Kammoun\,$^{7,8}$, 
Kunal Mooley\,$^{1}$, 
Megan Masterson\,$^{9}$, 
Lea Marcotulli\,$^{10,11,\dagger}$
Arne Rau\,$^{12}$,
George A. Younes,$^{6}$, 
Daniel Stern\,$^{13}$, 
Javier A. Garc\'ia\,$^{14,1}$ 
and Kristin Madsen$^{14}$}

\def\Address{
$^{1}$Cahill Center for Astrophysics, California Institute of Technology, 1216 East California Boulevard, Pasadena, CA 91125, USA \\
$^{2}$Department of Astronomy, University of California, Berkeley, CA 94720-3411, USA\\
$^{3}$Department of Physics, University of California, Berkeley, CA 94720-3411, USA\\
$^{4}$Department of Astronomy and Astrophysics, The University of Chicago, Chicago, IL 60637, USA\\
$^{5}$Research Center for the Early Universe, Graduate School of Science, University of Tokyo, Bunkyo, Tokyo 113-0033, Japan\\
$^{6}$Department of Physics, The George Washington University, 725 21st St. NW, Washington, DC 20052, USA \\
$^{7}$IRAP, Universit\'{e} de Toulouse, CNRS, UPS, CNES, 9, Avenue du Colonel Roche, BP 44346, F-31028, Toulouse Cedex 4, France \\
$^{8}$INAF -- Osservatorio Astrofisico di Arcetri, Largo Enrico Fermi 5, I-50125 Firenze, Italy \\
$^{9}$MIT Kavli Institute for Astrophysics and Space Research, Massachusetts Institute of Technology, Cambridge, MA 02139, USA\\
$^{10}$Yale Center for Astronomy \& Astrophysics, 52 Hillhouse Avenue, New Haven, CT 06511, USA \\
$^{11}$Department of Physics, Yale University, P.O. Box 208120, New Haven, CT 06520, USA\\
$\dagger$ NHFP Einstein Fellow\\
$^{12}$Max-Planck-Institut f\"{u}r extraterrestrische Physik, Giessenbachstrasse 1, 85748 Garching, Germany\\
$^{13}$Jet Propulsion Laboratory, California Institute of Technology, Pasadena, CA 91109, USA\\
$^{14}$X-Ray Astrophysics Laboratory, NASA Goddard Space Flight Center, Greenbelt, MD 20771, USA}

\def\corrAuthor{Murray Brightman}

\def\corrEmail{murray@srl.caltech.edu}

\begin{document}
\onecolumn
\firstpage{1}

\title{The High Energy X-ray Probe (\textit{HEX-P}): Sensitive broadband X-ray observations of transient phenomena in the 2030s} 

\author[\firstAuthorLast ]{\Authors} 
\address{} 
\correspondance{} 

\extraAuth{}

\maketitle

\begin{abstract}

\hexp\ will launch at a time when the sky is being routinely scanned for transient gravitational wave, electromagnetic and neutrino phenomena that will require the capabilities of a sensitive, broadband X-ray telescope for follow up studies. These include the merger of compact objects such as neutron stars and black holes, stellar explosions, and the birth of new compact objects. \hexp\ will probe the accretion and ejecta from these transient phenomena through the study of relativistic outflows and reprocessed emission, provide unique capabilities for understanding jet physics, and potentially revealing the nature of the central engine.

\tiny
 \keyFont{ \section{Keywords:} transients: neutron star mergers, transients: black hole - neutron star mergers, transients: black hole mergers, transients: supernovae, transients: fast blue optical transients} 
\end{abstract}

\section{Introduction}
X-ray transients comprise some of the most interesting and energetic events in the Universe, and are usually powered by the birth of, accretion onto, and mergers of compact objects such as black holes and neutron stars. Surveys for astrophysical transients are now possible through electromagnetic, gravitational wave, and neutrino searches. As these transient surveys become more prevalent and more sensitive, it will be essential to have the capability to study the X-ray emission from the transients they uncover over a wide a range in energy and fluxes in order to understand the physics at their core. 

Starting in the 2030s, the next generation of gravitational-wave observatories are expected to come online, namely the Einstein Telescope \citep{punturo10, maggiore20}, the Cosmic Explorer \citep{reitze19}, and the space-based {\it Laser Interferometer Space Antenna} \citep[{\it LISA};][]{amaro17}. These observatories will detect the mergers of neutron stars and black holes out to large cosmological distances with improved sky localizations that will significantly increase the chances to identify their electromagnetic counterparts. The future of optical transients will be dominated by the Vera Rubin Observatory and its large synoptic survey where 1--2 million alerts are expected per night, as well as one million new supernovae from Rubin's Legacy Survey of Space and Time \citep[LSST;][]{ivezic19}. A rich sample of transients at longer wavelengths will also be found by the next generation radio observatories, particularly by the Square-Kilometer Array \citep[SKA;][]{dewdney09} and its pathfinders which are already in operation, such as the Australian SKA Pathfinder \citep[ASKAP,][]{johnston08} and the South African MeerKAT (Jonas et al., 2016). In the UV, {\it ULTRASAT} is approved for launch and will deliver a large number of UV transients \citep{shvartzvald23}.  NASA is also currently concluding the down-select for the next Medium Explorer, both of which emphasize time-domain capabilities: either the {\it Ultraviolet Explorer} ({\it UVEX}) observing at UV energies providing large field of view UV photometry and rapid spectroscopy, or the {\it Survey and Time-domain Astrophysical Research eXplorer} ({\it STAR-X}) observing at soft X-ray and UV energies, is expected to launch at the end of this decade. Finally, the next generation of neutrino detectors, namely the KM3NeT in the Mediterranean Sea \citep{katz06}, the Gigaton Volume Detector in Lake Baikal \citep[Baikal-GVD;][]{belolaptikov97}, and IceCube-Gen2 at the South Pole \citep{icecube2} also promise to yield a rich new view of the high-energy Universe, including stellar explosions.

\cite{polzin22} presents a compilation of all known classes of X-ray transients to date, covering a wide range in type, flux, and timescale. These range from low-luminosity events typically seen in our Galaxy, such as X-ray binary outbursts, to high-luminosity events seen in other galaxies, such as tidal disruption events \citep[TDEs -- e.g.,][]{rees88, evans89, auchettl17,gezari21}. Both of these examples are produced by the accretion of stellar material onto a compact object such as a neutron star or a black hole. \cite{polzin22} also includes X-ray transients produced by the possible formation of a compact object, such as supernovae \citep[SNe -- e.g.,][]{chevalier17} and more recently discovered fast blue optical transients \citep[FBOTs -- e.g.,][]{drout14,arcavi16,ho21}, which might also be manifestations of (super-Eddington) accreting compact-objects.

\begin{figure}[h!]
\begin{center}
\includegraphics[trim=10 20 20 0, width=180mm]{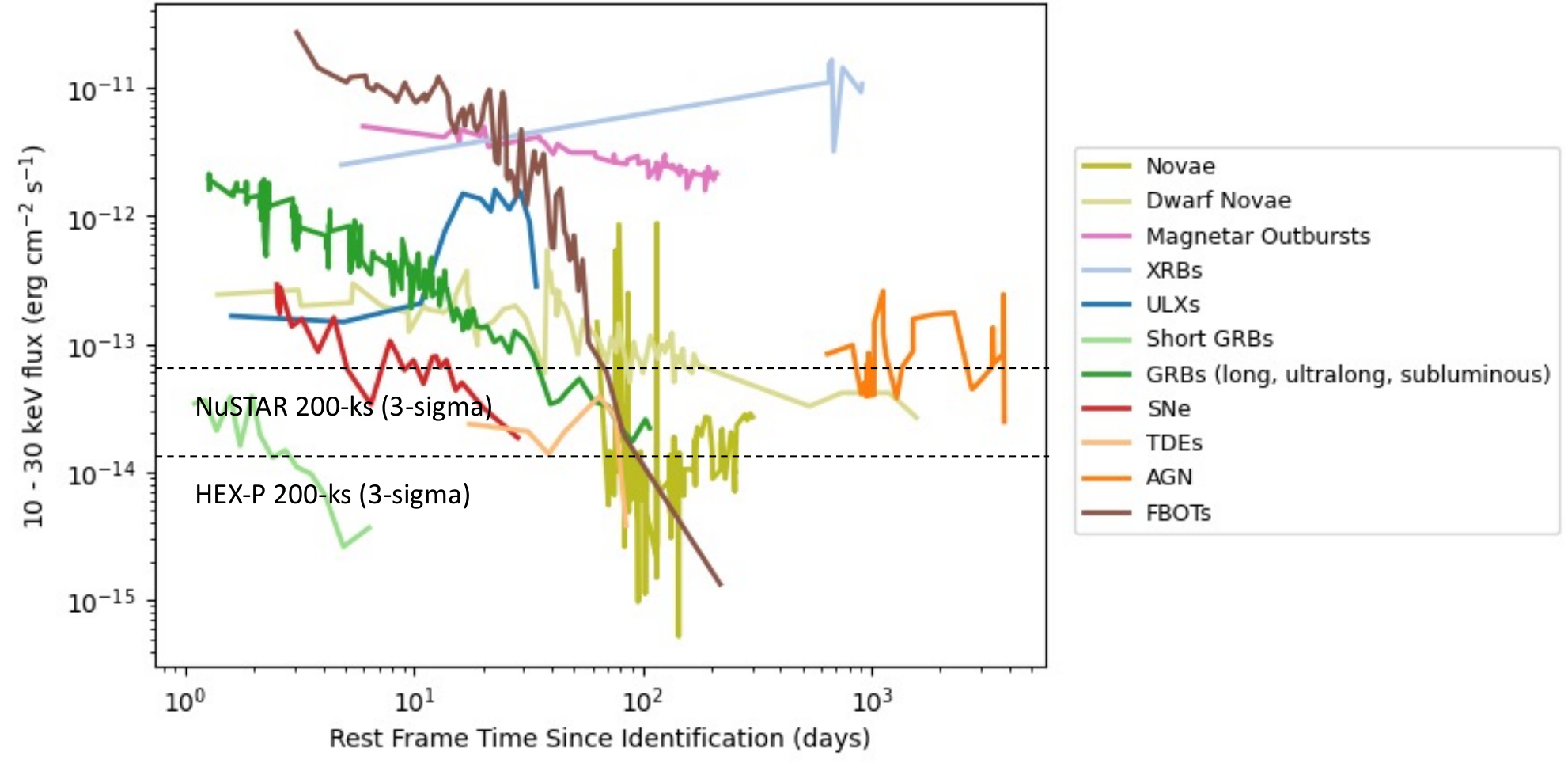}
\end{center}
\caption{10--30 keV X-ray lightcurves of known X-ray transients, adapted from \cite{polzin22}. \hexp\ will probe deeper in sensitivity than \nustar\ while matching \xmm\ and \chandra\ at lower energies. This will enable sensitive broadband studies of all known X-ray transient types from $\leq$ 1 day post identification.}
\label{fig_polzin}
\end{figure}

The Probe-class \hexp\ mission (Madsen et al., 2023) will provide sensitive broadband X-ray observations including roughly an order of magnitude improvement in the 10--30 keV band over \nustar, with significantly sharper angular resolution than either \xmm\ (below 10~keV) or \nustar\ (above 10~keV). This sensitivity over a broad X-ray band will revolutionize the field of high-energy transient science, as shown in Figure \ref{fig_polzin} \citep[adapted from][]{polzin22}, which shows that the faintest well-sampled lightcurves of each transient type to date will all be easily detected by \hexp\ in the 10--30 keV band at early times.

Below we describe the \hexp\ mission and the transient science for which we expect \hexp\ to provide invaluable insights. We begin with the electromagnetic counterparts to gravitational wave events, followed by FBOTs, and SNe. We will also describe how \hexp\ will yield insights into changing-look AGN (CLAGN), TDEs, blazars and fast radio bursts (FRBs) which are covered in more detail in separate papers (Kammoun et al., 2023, Marcotulli et al., 2023, Alford et al. 2023).

\section{Mission Design}
\label{sec_mission}

The {\it High-Energy X-ray Probe} (\hexp; Madsen et al., 2023) is a probe-class mission concept that offers sensitive broad-band X-ray coverage (0.2--80\,keV) with exceptional spectral, timing and angular capabilities.  It features two high-energy telescopes (HETs) that focus hard X-rays and one low-energy telescope (LET) that focuses lower-energy X-rays.

The LET consists of a segmented mirror assembly coated with Ir on monocrystalline silicon that achieves an angular resolution of $3.5''$, and a low-energy DEPFET detector, of the same type as the Wide Field Imager \citep[WFI;][]{meidinger20} onboard {\it Athena} \citep{nandra13}. It has $512 \times 512$ pixels that cover a field of view of $11.3'\times 11.3'$. The LET has an effective passband of 0.2--25\,keV, and a full frame readout time of 2\,ms, which can be operated in a 128 and 64 channel window mode for higher count-rates to mitigate pile-up and faster readout. Pile-up effects remain below an acceptable limit of ${\sim}1\%$ for fluxes up to ${\sim}100$\,mCrab in the smallest window configuration. Excising the core of the PSF, a common practice in X-ray astronomy, will allow for observations of brighter sources, with a typical loss of up to ${\sim}60\%$ of the total photon counts.

The HET consists of two co-aligned telescopes and detector modules. The optics are made of Ni-electroformed full shell mirror substrates, leveraging the heritage of \xmm\ \citep{Jansen2001}, and coated with Pt/C and W/Si multilayers for an effective passband of $2-80$\,keV. The high-energy detectors are of the same type and flown on \nustar\ \citep{Harrison2013}, and they consist of 16 CZT sensors per focal plane, tiled $4\times4$, for a total of $128\times128$ pixel spanning a field of view slightly larger than for the LET, of 13.4\arcmin$\times$13.4\arcmin.

The response time to target of opportunity (ToO) observation requests will be $<24$ hours enabling fast observations of transient events. As shown in Figure \ref{fig_polzin} this will be more than enough to observe and detect and characterize all known X-ray transients in the 10--30 keV band while they are still bright. Furthermore, \hexp's location at L1 will allow a large fraction of the sky to be observed at any one time - the field of regard will be 3$\pi$ steradians. This will also enable the fast observation of transient events.

The broad X-ray passband, superior sensitivity, fast response time and large field of regard will provide a unique opportunity to enable revolutionary new insights into the wealth of transient astrophysical phenomena that the facilities discussed above are expected to discover.

\section{Simulations}
\label{sec_sims}

All simulations presented here were produced with a set of response files that represents the observatory performance based on current best estimates as of Spring 2023 (see Madsen et al., 2023). The effective area is derived from raytracing calculations for the mirror design including obscuration by all known structures. The detector responses are based on simulations performed by the respective hardware groups, with an optical blocking filter for the LET and a Be window and thermal insulation for the HET. The LET background was derived from a GEANT4 simulation \citep{Eraerds2021} of the WFI instrument, and the HET background was derived from a GEANT4 simulation of the \nustar\ instrument. Both assume \hexp\ is in an L1 orbit. 

\section{New insights into transient phenomena by HEX-P}

\subsection{Probing the ejecta and remnant emission from the mergers of neutron stars and black holes detected in gravitational waves}
\label{sec_emgw}

With the detection of the first binary black hole merger in 2015 \citep{abbott16}, LIGO-Virgo opened a new gravitational wave window onto the Universe. This new era of multi-messenger astronomy with gravitational waves expanded with the LIGO-Virgo detection of GWs and photons from the binary neutron star merger GW170817 \citep{abbott17}, which yielded a wide range of scientific results, informing us about gravitational physics, nucleosynthesis, extreme states of nuclear matter, relativistic explosions and jets, and cosmology \citep[e.g.,][]{cowperthwaite17, margutti17, troja17, margutti18, mooley18, mooley18b, margutti21, mooley22}.

\begin{figure}[h!]
\begin{center}
\includegraphics[trim=10 20 20 0, width=150mm]{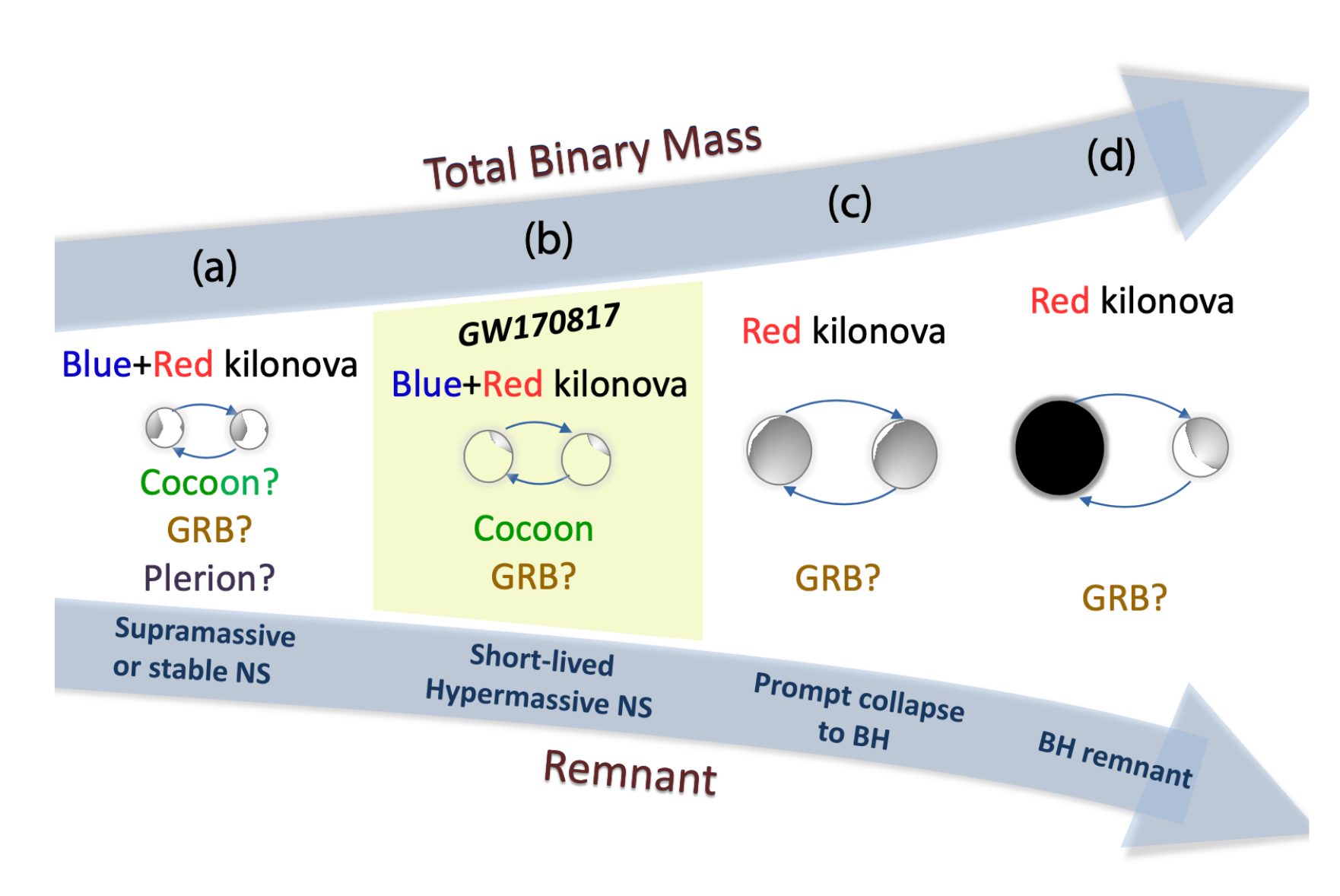}
\end{center}
\caption{The range of binary neutron star and neutron star-black hole merger events expected to produce an X-ray afterglow.  \hexp\ will provide a powerful tool for studying and understanding these extreme events.}
\label{fig_nsbh}
\end{figure}

The X-ray emission from GW170817 was produced by the collision between the fastest moving collimated ejecta and the circumburst medium \citep{margutti17, troja17, margutti18,davanzo18,hajela22}. GW170817 represents only an initial exploration of a rich scientific landscape populated by stellar evolution, explosions, and eventual mergers of massive binary systems, and there are still many unknowns regarding the outcome of binary neutron star and neutron star-black hole mergers. The total binary mass and binary mass ratio, as well as the nature of the final remnant, governs the mass, velocity, and direction of ejecta outflows and should lead to a diverse range of electromagnetic counterparts. For example, the properties of the associated explosion, known as a kilonova, is expected to depend on how long lived a hypermassive neutron star (HMNS) may last before collapsing to form a black hole (Figure \ref{fig_nsbh}). If black hole formation is prompt, a red kilonova that peaks in the near-infrared is expected, otherwise, a longer-lived HMNS could produce a bluer kilonova \citep[e.g.][]{metzger14}. The neutron star equation of state is also expected to affect the kilonova properties \citep{zhao23}.

\begin{figure}[h!]
\begin{center}
\includegraphics[trim=10 20 20 0, width=80mm]{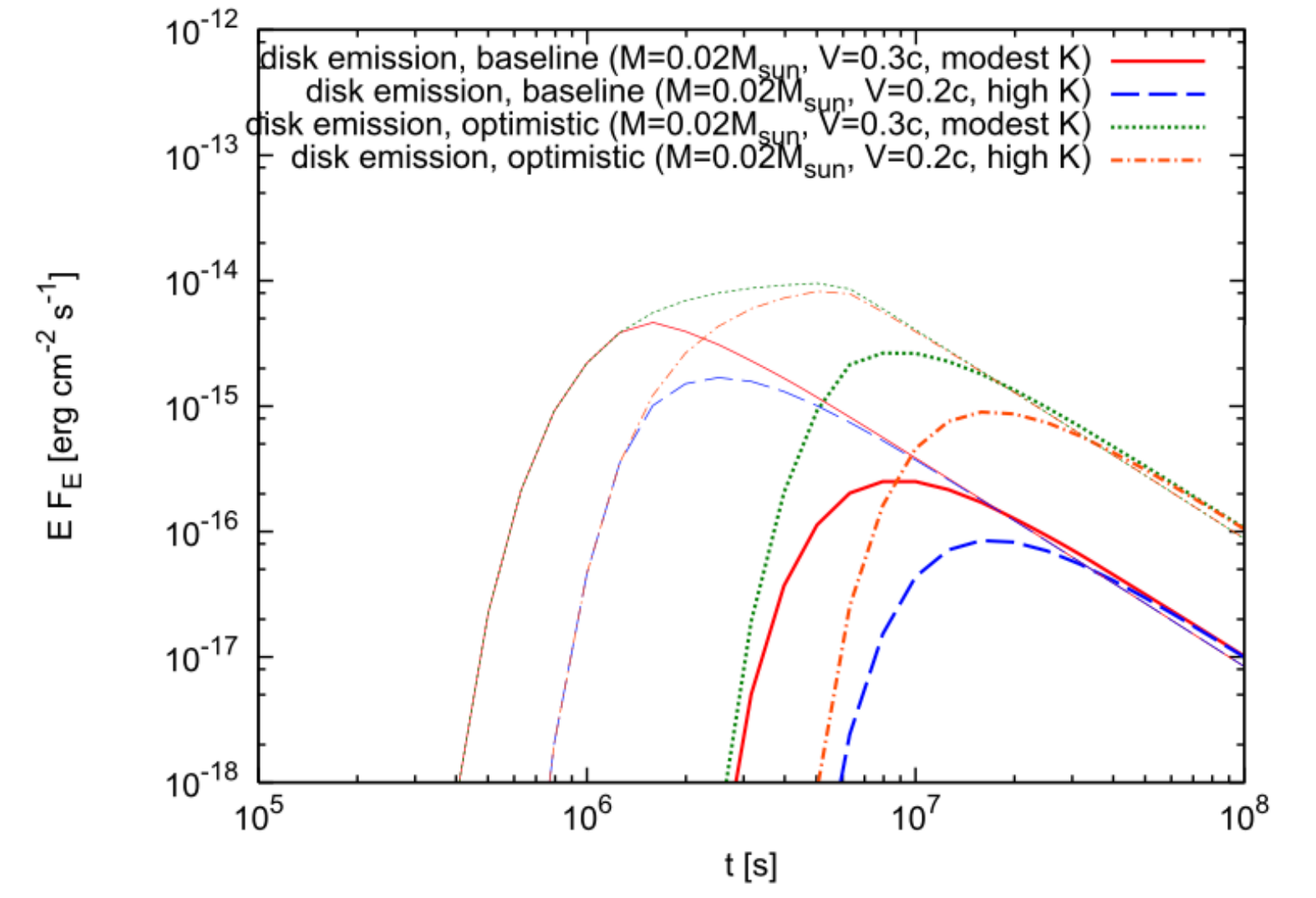}
\includegraphics[trim=10 20 20 0, width=80mm]{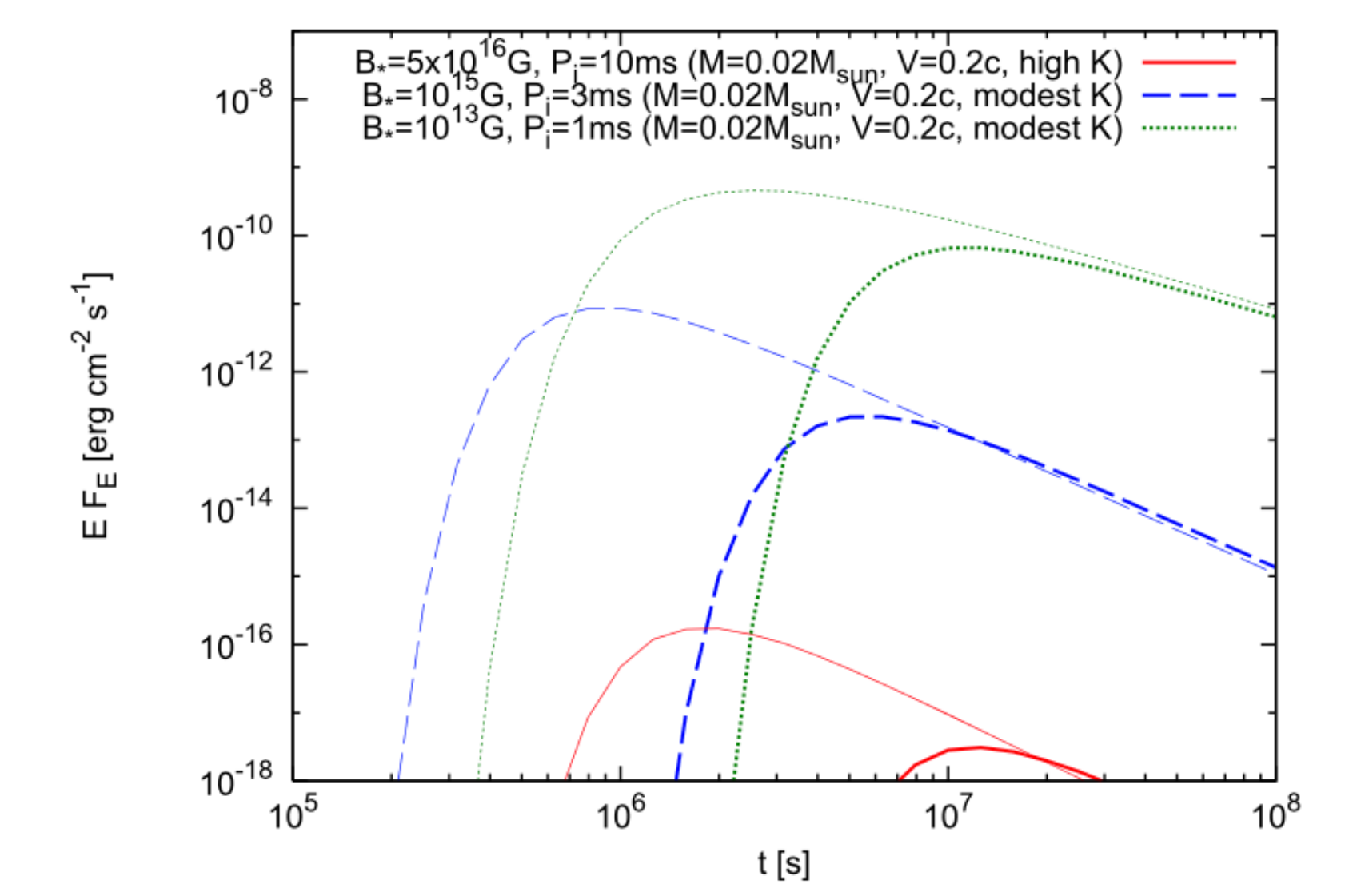}
\end{center}
\caption{Predicted X-ray lightcurves of emission from an accretion disk of a black hole remnant (left) and emission from a pulsar wind nebula of a neutron star remnant (right) taken from \cite{murase18}. In both cases, emission at 30 keV (thin lines) emerges prior to and is stronger than emission at 3 keV (thick lines) due to the the remnant being buried in the merger ejecta which is initially optically thick.}
\label{fig_murase}
\end{figure}

For binaries with significantly lower total mass than GW170817, does a long-lived supermassive or stable neutron star remnant form? \hexp\ has the potential to observe emission from the merger remnant, be it a black hole or neutron star. \cite{murase18} calculates the expected X-ray emission from the merger remnant, which, depending on the nature of the remnant, they predict would come from the disk of a newly formed black hole or the pulsar wind nebula of a newly formed neutron star. In both cases, the emission would be highly absorbed at early times due to the remnant being buried in the merger ejecta (Figure \ref{fig_murase}). Hard X-rays would thus be the first to emerge, followed by softer X-rays as the ejecta become optically thin. Indeed, in the recent example of SN 2023ixf in the nearby Pinwheel Galaxy (M101; $d$ = 6.9 Mpc), \cite{grefenstette23} report on \nustar\ observations a highly absorbed X-ray spectrum 4~days after the explosion.  Soft X-ray observations by \swift\ failed to detect the supernova at such early times. Naturally, an X-ray observatory with hard X-ray sensitivity such as \hexp\ will be well suited to detect this kind of early and evolving remnant emission.

Assuming a distance of 40 Mpc, \cite{murase18} predicts that the 30 keV emission from a black hole merger remnant would peak at $\sim 10^{-14}$ \ergcms\ approximately $\sim10^{6}-10^{7}$\,s post-merger (i.e., weeks to month timescales). \hexp\ could reach these sensitivities at 30 keV (3$\sigma$) in a 500 ks observation. The emission from a neutron star is expected to be significantly brighter and therefore detectable to larger distances and/or shorter exposures.

\begin{figure}[h!]
\begin{center}
\includegraphics[trim=10 20 20 0, width=180mm]{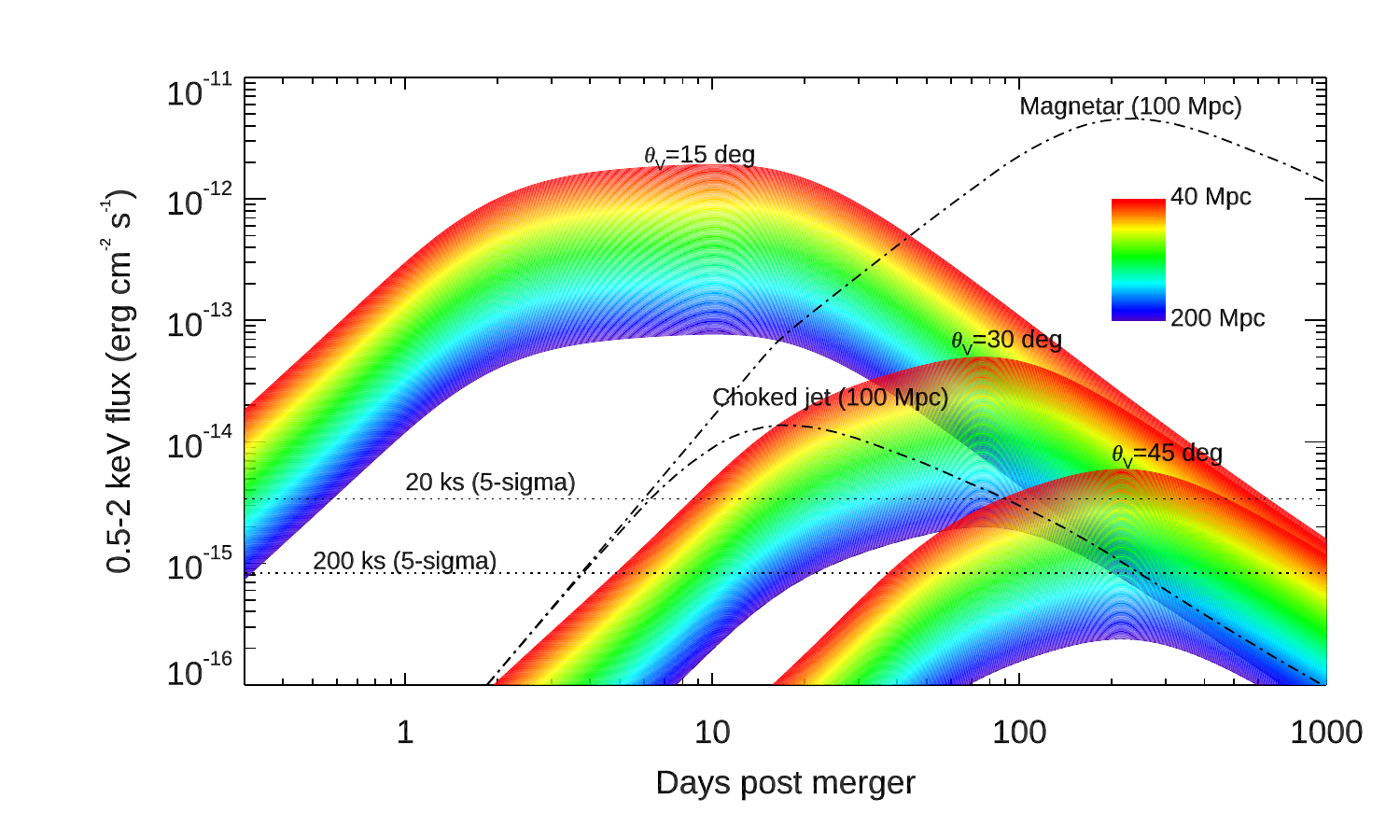}
\end{center}
\caption{X-ray lightcurves calculated from hydrodynamic simulations of a binary neutron star merger at distances of 40--200 Mpc and circumburst density of 0.1 cm$^{-3}$. Our simulation includes outflow from the jet interacting with the interstellar medium, with viewing angles of $\theta_{\rm V}=15$, 30 and 45 degrees. We also plot a prediction for a choked jet and the magnetar model. \hexp\ sensitivities are shown with dotted lines.}
\label{fig_nsns}
\end{figure}

X-ray lightcurves for events as bright as GW170817 provide a powerful means of probing ejecta properties such as energy and angular structure. From these bright events we can determine which binary neutron star mergers are capable of launching ultra-relativistic jets \citep[e.g.][]{nathanail21,sun22}. Potentially, all binary neutron star mergers generate ultrarelativistic jets.  If only some do, what determines the behavior? \hexp\ will be able to detect the afterglow emission from a neutron star-neutron star merger at distances up to 200 Mpc under a range of jet scenarios, and characterize its lightcurve across multiple energy bands, thereby answering these questions. We illustrate this in Figure \ref{fig_nsns} which shows lightcurves calculated from hydrodynamic simulations of a binary neutron star merger based on the semi-analytic code used in \cite{lazzati18}. The model has five free parameters: the viewing angle $\theta_{\rm v}$, the microphysical parameters $\epsilon_{\rm e}$ (the fraction of shock energy given to electrons) and $\epsilon_{\rm B}$ (the fraction of shock energy given to tangled magnetic field), the electrons population distribution index $p$, and the external medium density $n_{\rm ISM}$, which was assumed to be 0.1 cm$^{-3}$ and constant. The total kinetic energy of the fireball and its initial Lorentz factor, both dependent on the viewing angle, were taken from a hydrodynamic numerical simulation previously described in \cite{lazzati17}. Specifically, the energy of the blast wave, set by the numerical simulation, is $6\times10^{49}$ erg. Our simulation includes outflow from the jet interacting with the interstellar medium, with viewing angles of $\theta_{\rm V}=15$, 30 and 45 degrees. 

For a binary merger with significantly more mass than GW170817, we might expect a prompt collapse to a black hole. Although the formation of jets and cocoons is uncertain, the detection of the X-ray and radio afterglows may be able to disambiguate binary neutron star mergers from neutron star-black hole mergers of similar total mass. For neutron star-black hole mergers where the neutron star is disrupted beyond the innermost stable circular orbit, the tidal ejecta is expected to have much higher mass ($\sim0.1$\msol) than in a binary neutron star merger, and could yield a bright red kilonova. An absence of polar ejecta in this case would imply that jets, if launched, would escape to produce a short gamma-ray burst. However, there are large uncertainties and many unknowns in these predictions. \hexp, with spectral and timing capabilities over a wide bandpass, as well as $<24$ hr response time and good sky visibility, will be well positioned to observe the range of X-ray emission from the ejecta of neutron star-neutron star and neutron star-black hole mergers. 

\subsection{Black Hole-Black Hole Mergers in AGN Accretion Disks}
\label{sec_bbhs}

Binary black hole mergers come from two broad classes: a binary star evolutionary channel and a dynamical, or hierarchical, channel \citep[for a recent review, see][]{mapelli21}. Among dynamical mergers, sub-channels include mergers in globular clusters, mergers in quiescent galactic nuclei, and mergers in the accretion disks of active galactic nuclei (AGN). Since the most massive stars end their lived in pair instability supernovae which leave no compact remnant, the explosive deaths of massive stars are not thought capable of producing black holes in the ``upper mass gap'' range above $\sim 50M_{\odot}$.  Massive binary black hole merger progenitors above $\sim 100M_{\odot}$ therefore strongly imply a hierarchical merger origin, i.e., from consecutive merger events.  This, in turn, implies a dynamical origin, not a field binary origin. 

Since black holes can receive a strong kick at merger, mergers remnants are more easily retained in deep gravitational potentials, such as in the nuclei of galaxies. A promising location for hierarchical mergers are therefore AGN \citep{mckernan19}.  Merger kicks, even of large magnitude, are insufficient to escape an AGN environment, making AGN ideal for retaining and growing black holes via hierarchical mergers. AGN are expected to dominate the rate of mergers in the deep potential wells of galactic nuclei. Besides massive binary black hole mergers, other pointers to a significant contribution to binary black hole mergers from the AGN channel include highly asymmetric mass ratio black hole mergers and the observed anti-correlation between binary black hole mass ratio and effective spin \citep{callister21}, which at present can only be explained in the context of the AGN channel \citep{mckernan22}.

Unlike all other merger channels, detectable counterparts are expected from compact object mergers in AGN due to the baryon-rich, high density environment. Simple scaling arguments imply that the luminosity of the counterpart should scale roughly linearly with the binary total mass \citep{mckernan19}, implying more massive gravitational wave events create more luminous flares. Indeed, the first event with a candidate optical counterpart, reported by \citep{graham20}, was later identified as the most massive LIGO/Virgo merger observed to date, with a total mass of $150M_\odot$ \citep{abbott20}. \cite{kimura21} present a model of how compact object mergers in AGN accretion disks should appear, with outflows creating a bubble in the AGN accretion disk, and the merger remnant black hole subsequently recoiling into the dense AGN disk and producing strong X-ray outflow-breakout emission that can outshine the AGN. 

Though we are currently at very early stages, both theoretically and observationally, for understanding binary black hole mergers in AGN accretion disks, there are strong arguments to expect that this is an important gravitational wave merger channel, and that broadband X-ray emission will be a key tool for studying such events.  We also note that counterparts to black hole mergers in accretion disks provide a test of the dynamics of the merger and a probe of fundamental AGN disk properties \citep{vajpeyi22}. In addition, associated counterparts (i.e. spectroscopic redshifts) to gravitational wave mergers enable their use as a standard siren, providing a new, independent measurement of the Hubble constant \citep{chen22}.  \hexp\ is well posed to become an important tool for studying both binary black hole mergers in AGN accretion disks, as well as the larger population of gravitational wave merger events over a range of environments.

\subsection{Observing a potentially newly-formed compact object in fast blue optical transients}
\label{sec_fbots}

With the wealth of large area optical surveys such as the Zwicky Transient Facility (ZTF) and the upcoming Vera Rubin Observatory, a vast array of optical transients have been uncovered, with orders of magnitude more sources expected in the 2030s. Fast evolving luminous transients, or fast blue optical transients (FBOTs) are one such new class \citep{drout14,arcavi16,ho21}. Their fast evolution with rise times of $<10$ days and high luminosity in excess of $10^{43}$ \ergs\ imply a power source incompatible with the standard radioactive decay model for supernovae and can be explained by prolonged energy injection from a central compact object. Only a handful of such events have been discovered so far and are thought to comprise $<0.1$\% of the local supernova population \citep{ho23}.

The best studied FBOT to date is AT\,2018cow which, at a distance 60 Mpc, reached a peak X-ray luminosity of $\sim4\times10^{44}$ \ergs\ \citep{margutti19}. One of the defining properties of this event was its luminous and variable X-ray emission unprecedented among optical transients. A \nustar\ observation $\sim7.7$ days after discovery revealed an excess of hard X-ray emission above the soft X-ray power-law which is well modeled by reflection from Compton-thick material, i.e., the reprocessing of X-rays in an optically-thick medium such as an accretion disk.  Reflection spectra are characterized by the presence of Fe K fluorescence emission lines near 6.4-6.9~keV, an absorption Fe K-edge at $\sim$ 7-9~keV, and a broad featureless hump peaking at 20-30 keV produced by electron scattering. Modeling the high-energy spectrum of AT\,2018cow required aspherical ejecta. The reflection features faded rapidly and could not be constrained by \nustar\ in observations a few days later, precluding a detailed study of its nature. One possible explanation for the origin of the reflected component was a funnel formed by a super-Eddington accretion flow, is often used to explain the spectra of ultraluminous X-ray sources \citep{margutti19}. Similar funnel-like geometries for super-Eddington accretion have been invoked in TDEs too \citep[e.g.]{kara16,dai18}.

\begin{figure}[h!]
\begin{center}
\includegraphics[trim=10 20 20 0, width=85mm]{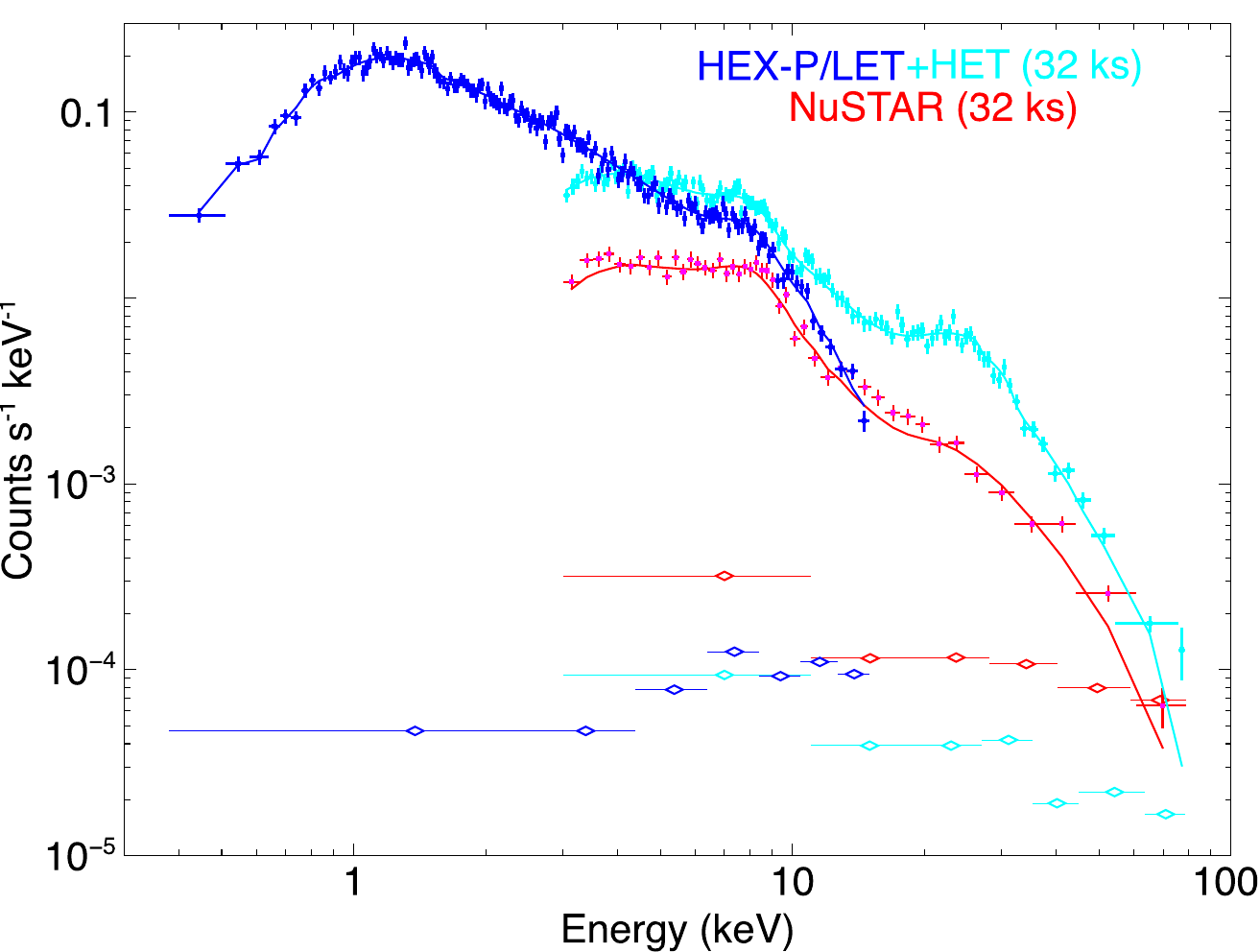}
\includegraphics[trim=10 20 20 0, width=85mm]{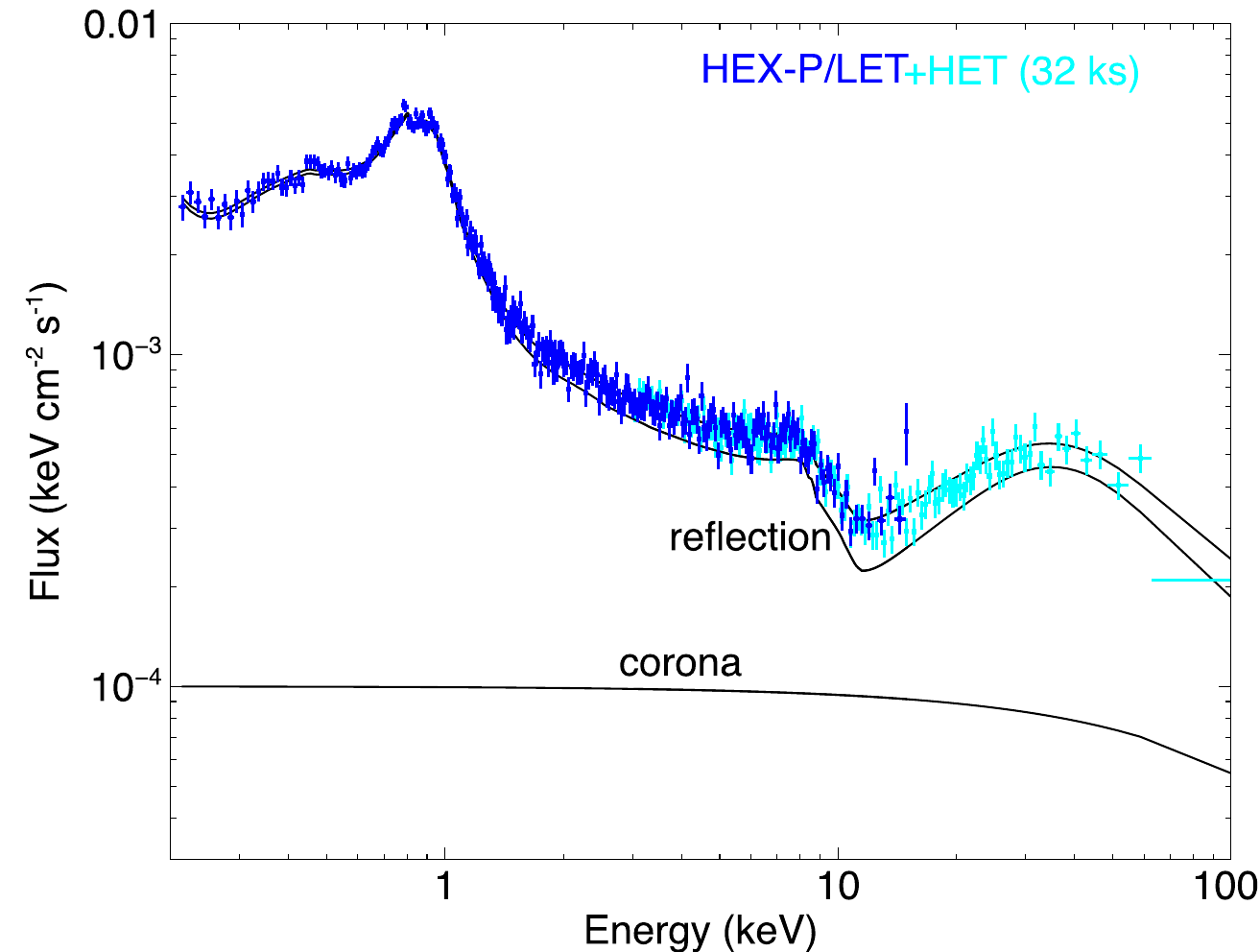}
\end{center}
\caption{Left - A simulated spectrum of \cow\ with \hexp\ compared to what \nustar\ observed (crosses). \hexp\ would have detected the transient over the 0.2--80 keV range, which combined with greater effective area and lower background (diamonds) than \nustar\ would have allowed a detailed study of its spectral components. Right - Unfolded \hexp\ spectrum of AT 2018cow showing the reflection and corona components.}
\label{fig_at2018cow}
\end{figure}

Due to its proximity and X-ray brightness, AT\,2018cow is the only transient to show such a Compton reflection feature so far. With the greater sensitivity and improved target-of-opportunity capabilities afforded by \hexp, studies of FBOTs will become more detailed and possible to larger distances and volume, allowing the determination of the central engine. We show a simulated \hexp\ spectrum of \cow\ in Figure \ref{fig_at2018cow}, compared to the \nustar\ actual data. We produced this by assuming the reflection+corona model ({\tt relxill+cutoffpl} in {\sc XSPEC}) described in \citep{margutti19} with parameters a=0.7, $\Gamma=1.0$, log$\xi=3.3$, $A_{\rm Fe}=4.9$, $E_{\rm cut}=170$ keV and $R_{\rm refl}=-12$ as shown in the right panel of Fig \ref{fig_at2018cow} and with the same 32-ks exposure time as the \nustar\ observation. Not only will \hexp\ provide the simultaneous soft X-ray data, it will provide a larger effective area and lower background than \nustar\ above 10 keV that will enable a more sensitive broadband X-ray spectral investigation of these sources, including detection of reflection components.

\hexp\ will also be able to detect AT\,2018cow-like events with a 10--30 keV signal-to-noise of 5 in a 20-ks observation up to 10 times further (600 Mpc), and in 1000 times the volume, which will undoubtedly lead to a leap in understanding of these events.

\subsection{Mapping the most extreme mass loss history of massive stars in interacting supernovae}
\label{sec_sne}

Along with possible engine-powered optical transients like \cow, a plethora of supernovae types will be detected by upcoming optical surveys. These include interacting supernovae where the ejecta from the stellar explosion interact with the ISM, producing high-temperature shocked emission that can reveal the mass loss history of the massive star \citep{chevalier17}. The high temperature of the shock can only be constrained with sensitive hard X-ray observations which have only been done with \nustar\ so far \citep[e.g. SNe 2010jli, 2014C and 2023ixf]{ofek14,chandra15,margutti17a,grefenstette23}. For the case of SN\,2014C at a distance of 14.7 Mpc, the shock produced bright hard X-ray emission that was observed with \chandra\ and \nustar\ \citep{margutti17a}. The hard X-ray coverage allowed for the measurement of the absorption and shock temperature (10--20 keV) as a function of time, which has only been achieved for SN\,2014C so far. These data constrain the density profile of the environment of the supernova. Recently, \nustar\ observations of SN 2023ixf from 4 days post-explosion have provided the earliest constraints on the absorption and shock temperature to date \citep{grefenstette23} and follow up observation promise to reveal more about the mass-loss history of its progenitor (Margutti et al. in prep.). 

\begin{figure}[h!]
\begin{center}
\includegraphics[trim=10 20 20 0, width=120mm]{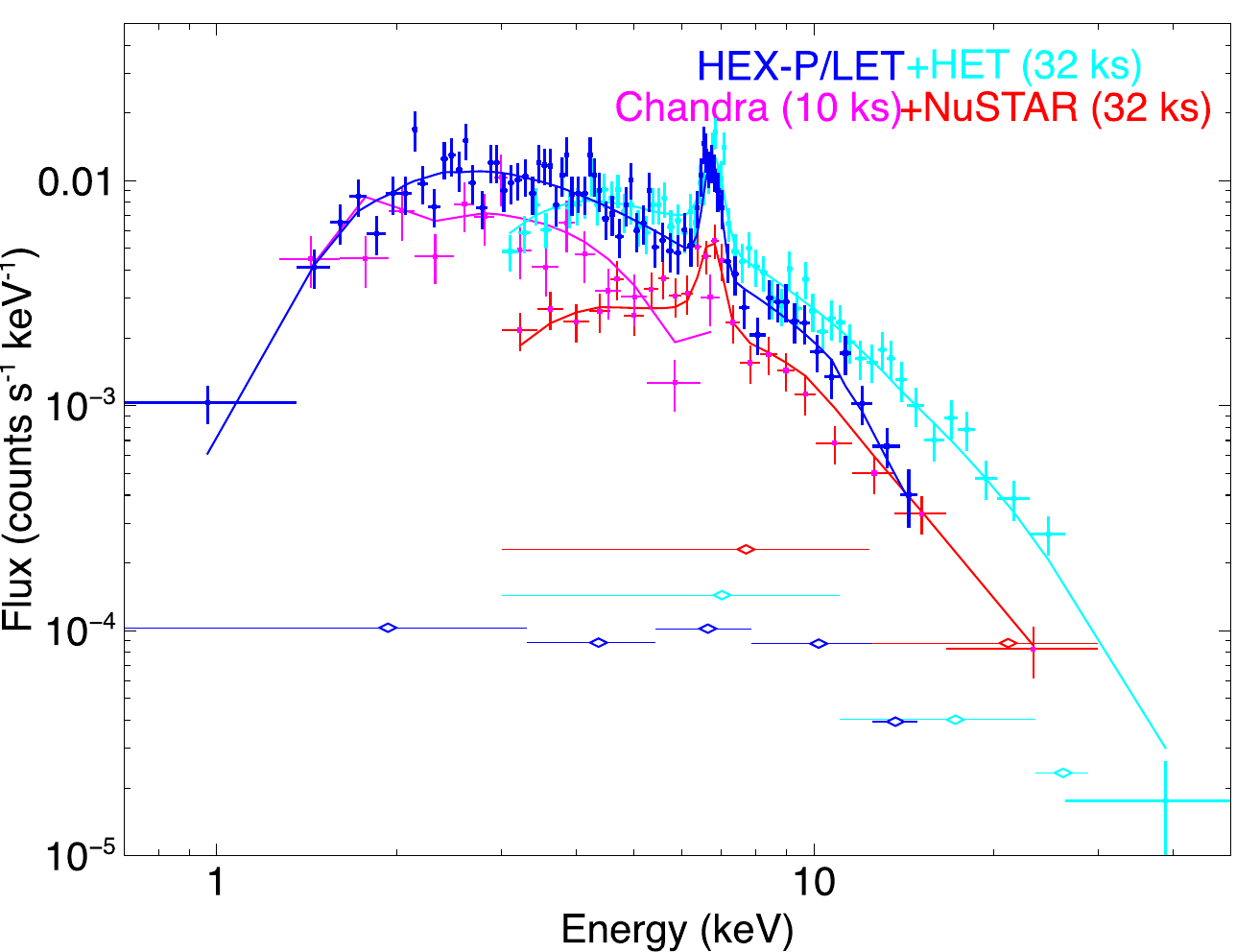}
\end{center}
\caption{A simulated spectrum of SN\,2014C with \hexp\ compared to what \chandra\ and \nustar\ observed (crosses). \hexp\ would have provided simultaneous data over the 0.2--50 keV range, which combined with greater effective area and lower background than \nustar\ (diamonds) would have provided a more detailed constraint on the shock properties.}
\label{fig_sn2014c}
\end{figure}

We show a simulated \hexp\ spectrum of SN\,2014C in Figure \ref{fig_sn2014c}. We produce these spectra by assuming the absorbed thermal bremsstrahlung with Gaussian line component spectral model described in \cite{margutti17a} ({\tt tbabs*bremss+gaussian} in {\sc xspec}) with parameters \nh$=3\times10^{22}$ \cmsq\ and $kT=18$ keV with a 0.3--10 keV flux of $10^{-12}$ \ergcms\ and the same 32-ks exposure time as \nustar. Compared to \chandra\ and \nustar, \hexp\ would have yielded a spectrum with signal-to-noise twice that of \nustar\ in the 3--30 keV band, providing better constraints on the shock temperature and absorption, where the uncertainties would be reduced by more than 50\%. More importantly, \hexp\ will also be able to detect SN\,2014C-like events with a 10--30 keV signal-to-noise of 5 in a 20-ks observation up to 3 times further (45 Mpc), and in 27 times the volume.

\subsection{Changing-look AGN and tidal disruption events}
\label{sec_tdes}

The advance of time-domain surveys is also revealing extreme variability in the accretion process onto supermassive black holes. Transient accretion events provide unique windows into the formation and evolution of accretion disks, the heating of the X-ray corona, and the connection between accretion and ejection via winds and jets. The X-ray band, in particular, is crucial to probing the innermost portions of the accretion flow, where strong outflows are launched and much of the gravitational energy is released. 

Recently there has been a class of AGN discovered that undergo rapid transitions on timescales of months to years, in which broad lines either appear or disappear in the optical spectrum. These sources are known as changing-look AGN \citep[CLAGN; for a recent review, see][]{ricci22}, and they challenge the classic unified model for AGN, whereby the presence of broad emission lines is solely a function of viewing angle. A particularly extreme source and the first CLAGN that was caught in the act of changing states, 1ES\,1927+654 showed dramatic and unprecedented X-ray variability as optical broad lines formed \citep{trakhtenbrot19,ricci20}. Shortly after the optical outburst, the canonical X-ray corona was destroyed \citep{ricci20}. Extensive X-ray monitoring revealed its reappearance, allowing for one of the first studies of how an AGN corona is formed \citep{ricci21,masterson22}. This dramatic behavior has been explained by either a TDE occurring in an AGN \citep{ricci20} or an inversion in the magnetic flux polarity \citep{scepi21,laha22}. \nustar\ observations were crucial for disentangling the soft photon index and low cutoff energy as the corona formed, while softer instruments like \xmm\ and \nicer\ probed the evolution of the inner accretion flow in 1ES\,1927+654, including rapid variability of the soft X-ray flux and a relativistic outflow launched from the inner disk \citep{masterson22}. Thus, broad energy coverage in the X-ray band is crucial to fully understanding this and other CLAGN. With its 0.2-80 keV bandpass, \hexp\ will be uniquely situated to probe the X-ray evolution, variability, and coupling between the inner disk and the corona in CLAGN. 

TDEs also hold key insights into how the X-ray corona is powered and formed, as numerous TDEs have shown spectral hardening and the formation of a corona at late-times \citep[$\sim$ a few years after the initial disruption; e.g.,][]{wevers21,yao22,guolo23}. They can also serve as probes of extreme accretion, as dramatic X-ray spectral variability, similar to that of 1ES\,1927+654, has also been seen in TDE candidates \citep[e.g., a repeating TDE discovered by eROSITA, which shows a repetitive creation and collapse of the X-ray corona;][]{liu23}. Likewise, as TDEs can also transition from super- to sub-Eddington as the mass fallback rate drops, \hexp\ can also be used to study the disk-corona connection during these transitions, thereby allowing for comparisons to state transitions in black hole binaries \citep{remillard06} and tests of the scale-invariant nature of accretion. Moreover, X-ray observations of TDEs can help illuminate the underlying accretion disk evolution, which is especially hard to probe with optical emission alone. Theoretical studies of TDEs have proposed that the apparent X-ray/optical dichotomy among TDEs is a result of viewing angle dependence \citep{dai18}, which has been recently supported by \swift\ and \xmm\ observations of numerous optically-selected TDEs \citep[e.g.][]{guolo23}. \hexp\ will be able to provide frequent and systematic X-ray follow-up of optically-selected TDEs, which are expected to become far more numerous in the next decade \citep[e.g.,][]{bricman20}. This is vital for understanding sample biases, as well as for probing other interesting accretion phenomena, like the evolution of outflows in TDEs. Finally, TDEs are also powerful probes of relativistic and multi-messenger astrophysics, with powerful on-axis jets detected in a few sources \citep{burrows11,cenko12,andreoni22,pasham23} and recent discoveries of high-energy neutrinos coincident with three optical TDE candidates \citep[contributing a sizeable fraction of the astrophysical high-energy neutrino flux, despite being found in non-jetted systems; e.g.,][]{stein21,vanvelzen21,reusch22}. Probing future jetted and neutrino-emitting TDEs in the X-ray band with \hexp\ will help to disentangle particle acceleration mechanisms and models for both jet and neutrino production.

\hexp\ will make significant contributions to our understanding of extreme accretion physics in CLAGN and TDEs, including improving key constraints on the evolution of coronal parameters like $\Gamma$ and $E_\mathrm{cut}$ by a factor of $\sim2$ with respect to \xmm\ and \nustar\ (Kammoun et al. 2023). More details on the coronal physics that we can glean from these transients with \hexp, as well as SMBH coronal physics in general, are described in Kammoun et al. (2023). Separately, Connors et al. (2023) describe the coronal physics of accreting stellar-mass BHs.

\subsection{Neutrinos and cosmic rays from blazars}
\label{sec_blazars}

Many AGN are known to launch relativistic jets which can extend up to kiloparsecs in length. These jets emit across the electromagnetic spectrum (radio to $\gamma$-ray) and are also multi-messenger (neutrino) emitters. When the viewing angle is close to edge on (viewing angles, $\theta_{\rm V}<5^{\circ}-10^{\circ}$) we observe a blazar, while at larger viewing angles we classify them as misaligned AGNs or radio galaxies \citep[e.g.,][]{urry95,hovatta09}.

Blazars are believed to be able to accelerate protons to very high energies. This was confirmed by the association of the $\gamma$-ray flaring blazar TXS~0506+056 with the IceCube neutrino event IceCube-170922A \citep{icecube18}. In order to probe the transition between the synchrotron and high-energy emission part of the spectrum, where signatures of electromagnetic cascades associated with the neutrino-producing photo-hadronic interactions should emerge \citep{murase18a} simultaneous broadband coverage from X-ray to MeV $\gamma$-rays is required, and has so far been lacking. This was shown in previous studies of a previous neutrino flare potentially linked to TXS~0506+056 where the lack of sensitive simultaneous broad-band X-ray spectral and temporal coverage did not allow for any conclusion on the emission mechanism \citep[e.g.][]{reimer19, zhang20}. Simultaneous broadband coverage of neutrino alerts in the direction of blazar sources could be the tipping point for jet models as well as confirming AGN as one of the sources of high-energy cosmic rays. 

For the first time \hexp\ will allow us to detect the spectral signatures that will conclusively yield the preferred high-energy emission mechanisms in jets, and, by extension, their composition. Furthermore, combined with current and upcoming observatories such as COSI, IXPE, CTA, IceCube Gen 2, \hexp\ will reveal the preferred particle acceleration and  emission mechanism of the brightest objects in the Universe, fulfilling the promises of multi-wavelength and multi-messenger science. The contribution to blazar science by \hexp\ is described in more detail in Marcotulli et al. (2023).

\subsection{Fast Radio Bursts}
\label{sec_frbs}

Fast Radio Bursts (FRBs) are millisecond duration transients discovered through their luminous sweeps in multi-band radio detectors \citep{petroff19}. Initially, FRBs were thought of as non-recurrent phenomena. However in the last decade advent of radio facilities such as MeerKAT, MeerTRAP and CHIME which could regularly monitor the known population of FRBs have led to a deluge of fast radio burst transients spanning a wide range of the parameter space. While most of the $\sim10^3$ FRBs detected so far do not seem to repeat yet, 24 of them have occasionally shown high burst rates clustered over a short duration. The emission models for both non-repeating and repeating FRBs are not yet well-constrained. 

In a breakthrough discovery, observational evidence for the magnetar model as a source of FRBs occurred on 2020 April 28, when an FRB-like radio burst was detected from the Galactic magnetar SGR 1935+2154 \citep{chime20, bochenek20}, in the winding hours of a major burst storm \citep{younes20}; it had a fluence rivaling those of the faint end of extragalactic FRBs. Moreover, the FRB occurred simultaneously with a bright, short X-ray burst, hence, relating it to magnetar activity and providing crucial evidence for its triggering mechanism \citep[e.g.,][]{mereghetti20,li21,ridnaia21}.

The discovery of FRB-like bursts from magnetars opened up new avenues for the study of extragalactic FRBs \citep[e.g.,][]{wadiasingh19}. This discovery opened up new questions such as (1) What is unique about the FRB-associated X-ray bursts, and why do the majority of X-ray bursts lack a radio counterpart? (2) What is the distribution of the spectral properties for FRB-associated X-ray bursts? Are their distinctive spectral properties universal across radio fluence? (3) Is the radio to X-ray flux ratio ($L_{\rm R}/L_{\rm X}$) constant for all FRB-like radio bursts? (4) What is the radio-X-ray time-lag across burst fluence? Answering these questions will require (1) a broad X-ray coverage given that the spectral energy distribution of magnetar burst peaks in the 20-30 keV range, (2) high timing resolution ($\lesssim1$~ms) for accurately measuring the radio-X-ray lag in burst arrival time \citep{mereghetti20}, (3) sensitivity to faint X-ray bursts, i.e., fluence $<10^{-7}$~erg~cm$^{-2}$, to sample a large fraction of the X-ray and radio burst fluence distribution given their steep shapes ($N\propto S^{-0.6}$, \citealt{younes20}). \hexp\ is the only facility to satisfy all the above criteria. The contribution to FRB science by \hexp\ is described in detail in Alford et al. (2023).

\section{Conclusions}
The 2030s will be an exciting era for time domain and multi-messenger astronomy, with a veritable orchestra of new facilities providing new views of new events in the universe. We are virtually guaranteed to discover events that are barely dreamed of currently.  Given this evolving landscape, it is essential that a sensitive, versatile, and flexible X-ray observatory will be available in the 2030s, able to do broadband X-ray studies, including imaging, spectroscopy, and timing, to understand the nature and the physics of these events.  From the early days of X-ray astronomy, it was revealed that the high-energy sky is highly variable -- and that X-rays provide a powerful and essential tool for understanding the range of phenomena.  This paper describes how \hexp\ will provide that needed capability for the 2030s.

\section*{Author Contributions}

M.B.\ is responsible for the creation of the manuscript, contributed to the science case and authored Sections \ref{sec_emgw},\ref{sec_fbots},\ref{sec_sne}, and made Figures \ref{fig_nsns}, \ref{fig_at2018cow} and \ref{fig_sn2014c}.
R.M.\ contributed to the science cases described in Sections \ref{sec_emgw},\ref{sec_fbots},\ref{sec_sne}. 
A.P.\ compiled Figure \ref{fig_polzin}.
A.J.\ contributed to the science case described in Section \ref{sec_emgw}.
K.H.\ contributed to the science case described in Section \ref{sec_emgw} and produced the data for Figure \ref{fig_nsns}.
J.A.J.A.\  contributed to the science case and authored Section \ref{sec_frbs}.
E. K.\  contributed to the science case and authored Section \ref{sec_tdes}.
K. M.\  contributed to the science case and authored Section \ref{sec_emgw}.
G. H.\  contributed to the science case and authored Section \ref{sec_emgw}.
M. M.\  contributed to the science case and authored Section \ref{sec_tdes}.
L. M.\  contributed to the science case and authored Section \ref{sec_blazars}.
A. R.\  contributed to the science case and authored Section \ref{sec_blazars}.
G. Y.\  contributed to the science case and authored Section \ref{sec_frbs}.
D. S.\  contributed to the science case and authored Sections \ref{sec_bbhs} and \ref{sec_mission}.
J. G.\  contributed to the science case and authored Section \ref{sec_mission}.
K. M.\  contributed to the science case and authored Section \ref{sec_mission}.
All authors contributed to the final editing of the manuscript.

\section*{Funding}

The work of D.S.\ was carried out at the Jet Propulsion Laboratory, California Institute of Technology, under a contract with NASA. E.K. acknowledges financial support from the Centre National d?Etudes Spatiales (CNES). G.H. and K.M. acknowledge support from the National Science Foundation Grant AST-1911199.

\bibliographystyle{Frontiers-Harvard}
\bibliography{HEX-P_transients_paper.bbl}

\begin{thebibliography}{95}
\providecommand{\natexlab}[1]{#1}
\expandafter\ifx\csname urlstyle\endcsname\relax
  \providecommand{\doi}[1]{doi:\discretionary{}{}{}#1}\else
  \providecommand{\doi}{doi:\discretionary{}{}{}\begingroup
  \urlstyle{rm}\Url}\fi
\providecommand{\selectlanguage}[1]{\relax}
\providecommand{\bibAnnoteFile}[1]{%
  \IfFileExists{#1}{\begin{quotation}\noindent\textsc{Key:} #1\\
  \textsc{Annotation:}\ \input{#1}\end{quotation}}{}}
\providecommand{\bibAnnote}[2]{%
  \begin{quotation}\noindent\textsc{Key:} #1\\
  \textsc{Annotation:}\ #2\end{quotation}}

\bibitem[{Abbott et~al.(2016)Abbott, Abbott, Abbott, Abernathy, Acernese,
  Ackley et~al.}]{abbott16}
Abbott, B.~P., Abbott, R., Abbott, T.~D., Abernathy, M.~R., Acernese, F.,
  Ackley, K., et~al. (2016).
\newblock Observation of gravitational waves from a binary black hole merger.
\newblock \emph{Phys. Rev. Lett.} 116, 061102.
\newblock \doi{10.1103/PhysRevLett.116.061102}
\bibAnnoteFile{abbott16}

\bibitem[{{Abbott} et~al.(2017){Abbott}, {Abbott}, {Abbott}, {Acernese},
  {Ackley}, {Adams} et~al.}]{abbott17}
{Abbott}, B.~P., {Abbott}, R., {Abbott}, T.~D., {Acernese}, F., {Ackley}, K.,
  {Adams}, C., et~al. (2017).
\newblock {GW170817: Observation of Gravitational Waves from a Binary Neutron
  Star Inspiral}.
\newblock \emph{Physical Review Letters} 119, 161101.
\newblock \doi{10.1103/PhysRevLett.119.161101}
\bibAnnoteFile{abbott17}

\bibitem[{{Abbott} et~al.(2020){Abbott}, {Abbott}, {Abraham}, {Acernese},
  {Ackley}, {Adams} et~al.}]{abbott20}
{Abbott}, R., {Abbott}, T.~D., {Abraham}, S., {Acernese}, F., {Ackley}, K.,
  {Adams}, C., et~al. (2020).
\newblock {GW190521: A Binary Black Hole Merger with a Total Mass of 150
  M$_{{\ensuremath{\odot}}}$}.
\newblock \emph{\prl} 125, 101102.
\newblock \doi{10.1103/PhysRevLett.125.101102}
\bibAnnoteFile{abbott20}

\bibitem[{{Amaro-Seoane} et~al.(2017){Amaro-Seoane}, {Audley}, {Babak},
  {Baker}, {Barausse}, {Bender} et~al.}]{amaro17}
{Amaro-Seoane}, P., {Audley}, H., {Babak}, S., {Baker}, J., {Barausse}, E.,
  {Bender}, P., et~al. (2017).
\newblock {Laser Interferometer Space Antenna}.
\newblock \emph{arXiv e-prints} ,
  arXiv:1702.00786\doi{10.48550/arXiv.1702.00786}
\bibAnnoteFile{amaro17}

\bibitem[{{Andreoni} et~al.(2022){Andreoni}, {Coughlin}, {Perley}, {Yao}, {Lu},
  {Cenko} et~al.}]{andreoni22}
{Andreoni}, I., {Coughlin}, M.~W., {Perley}, D.~A., {Yao}, Y., {Lu}, W.,
  {Cenko}, S.~B., et~al. (2022).
\newblock {A very luminous jet from the disruption of a star by a massive black
  hole}.
\newblock \emph{\nat} 612, 430--434.
\newblock \doi{10.1038/s41586-022-05465-8}
\bibAnnoteFile{andreoni22}

\bibitem[{{Arcavi} et~al.(2016){Arcavi}, {Wolf}, {Howell}, {Bildsten},
  {Leloudas}, {Hardin} et~al.}]{arcavi16}
{Arcavi}, I., {Wolf}, W.~M., {Howell}, D.~A., {Bildsten}, L., {Leloudas}, G.,
  {Hardin}, D., et~al. (2016).
\newblock {Rapidly Rising Transients in the Supernova{\textemdash}Superluminous
  Supernova Gap}.
\newblock \emph{\apj} 819, 35.
\newblock \doi{10.3847/0004-637X/819/1/35}
\bibAnnoteFile{arcavi16}

\bibitem[{{Auchettl} et~al.(2017){Auchettl}, {Guillochon}, and
  {Ramirez-Ruiz}}]{auchettl17}
{Auchettl}, K., {Guillochon}, J., and {Ramirez-Ruiz}, E. (2017).
\newblock {New Physical Insights about Tidal Disruption Events from a
  Comprehensive Observational Inventory at X-Ray Wavelengths}.
\newblock \emph{\apj} 838, 149.
\newblock \doi{10.3847/1538-4357/aa633b}
\bibAnnoteFile{auchettl17}

\bibitem[{{Belolaptikov} et~al.(1997){Belolaptikov}, {Bezrukov}, {Borisovets},
  {Budnev}, {Bugaev}, {Chensky} et~al.}]{belolaptikov97}
{Belolaptikov}, I.~A., {Bezrukov}, L.~B., {Borisovets}, B.~A., {Budnev}, N.~M.,
  {Bugaev}, E.~V., {Chensky}, A.~G., et~al. (1997).
\newblock {The Baikal underwater neutrino telescope: Design, performance, and
  first results}.
\newblock \emph{Astroparticle Physics} 7, 263--282.
\newblock \doi{10.1016/S0927-6505(97)00022-4}
\bibAnnoteFile{belolaptikov97}

\bibitem[{{Bochenek} et~al.(2020){Bochenek}, {Ravi}, {Belov}, {Hallinan},
  {Kocz}, {Kulkarni} et~al.}]{bochenek20}
{Bochenek}, C.~D., {Ravi}, V., {Belov}, K.~V., {Hallinan}, G., {Kocz}, J.,
  {Kulkarni}, S.~R., et~al. (2020).
\newblock {A fast radio burst associated with a Galactic magnetar}.
\newblock \emph{\nat} 587, 59--62.
\newblock \doi{10.1038/s41586-020-2872-x}
\bibAnnoteFile{bochenek20}

\bibitem[{{Bricman} and {Gomboc}(2020)}]{bricman20}
{Bricman}, K. and {Gomboc}, A. (2020).
\newblock {The Prospects of Observing Tidal Disruption Events with the Large
  Synoptic Survey Telescope}.
\newblock \emph{\apj} 890, 73.
\newblock \doi{10.3847/1538-4357/ab6989}
\bibAnnoteFile{bricman20}

\bibitem[{{Burrows} et~al.(2011){Burrows}, {Kennea}, {Ghisellini}, {Mangano},
  {Zhang}, {Page} et~al.}]{burrows11}
{Burrows}, D.~N., {Kennea}, J.~A., {Ghisellini}, G., {Mangano}, V., {Zhang},
  B., {Page}, K.~L., et~al. (2011).
\newblock {Relativistic jet activity from the tidal disruption of a star by a
  massive black hole}.
\newblock \emph{\nat} 476, 421--424.
\newblock \doi{10.1038/nature10374}
\bibAnnoteFile{burrows11}

\bibitem[{{Callister} et~al.(2021){Callister}, {Haster}, {Ng}, {Vitale}, and
  {Farr}}]{callister21}
{Callister}, T.~A., {Haster}, C.-J., {Ng}, K. K.~Y., {Vitale}, S., and {Farr},
  W.~M. (2021).
\newblock {Who Ordered That? Unequal-mass Binary Black Hole Mergers Have Larger
  Effective Spins}.
\newblock \emph{\apjl} 922, L5.
\newblock \doi{10.3847/2041-8213/ac2ccc}
\bibAnnoteFile{callister21}

\bibitem[{{Cenko} et~al.(2012){Cenko}, {Krimm}, {Horesh}, {Rau}, {Frail},
  {Kennea} et~al.}]{cenko12}
{Cenko}, S.~B., {Krimm}, H.~A., {Horesh}, A., {Rau}, A., {Frail}, D.~A.,
  {Kennea}, J.~A., et~al. (2012).
\newblock {Swift J2058.4+0516: Discovery of a Possible Second Relativistic
  Tidal Disruption Flare?}
\newblock \emph{\apj} 753, 77.
\newblock \doi{10.1088/0004-637X/753/1/77}
\bibAnnoteFile{cenko12}

\bibitem[{{Chandra} et~al.(2015){Chandra}, {Chevalier}, {Chugai}, {Fransson},
  and {Soderberg}}]{chandra15}
{Chandra}, P., {Chevalier}, R.~A., {Chugai}, N., {Fransson}, C., and
  {Soderberg}, A.~M. (2015).
\newblock {X-Ray and Radio Emission from Type IIn Supernova SN 2010jl}.
\newblock \emph{\apj} 810, 32.
\newblock \doi{10.1088/0004-637X/810/1/32}
\bibAnnoteFile{chandra15}

\bibitem[{{Chen} et~al.(2022){Chen}, {Haster}, {Vitale}, {Farr}, and
  {Isi}}]{chen22}
{Chen}, H.-Y., {Haster}, C.-J., {Vitale}, S., {Farr}, W.~M., and {Isi}, M.
  (2022).
\newblock {A standard siren cosmological measurement from the potential
  GW190521 electromagnetic counterpart ZTF19abanrhr}.
\newblock \emph{\mnras} 513, 2152--2157.
\newblock \doi{10.1093/mnras/stac989}
\bibAnnoteFile{chen22}

\bibitem[{{Chevalier} and {Fransson}(2017)}]{chevalier17}
{Chevalier}, R.~A. and {Fransson}, C. (2017).
\newblock {Thermal and Non-thermal Emission from Circumstellar Interaction}.
\newblock In \emph{Handbook of Supernovae}, eds. A.~W. {Alsabti} and
  P.~{Murdin}. 875.
\newblock \doi{10.1007/978-3-319-21846-5_34}
\bibAnnoteFile{chevalier17}

\bibitem[{{CHIME/FRB Collaboration} et~al.(2020){CHIME/FRB Collaboration},
  {Andersen}, {Bandura}, {Bhardwaj}, {Bij}, {Boyce} et~al.}]{chime20}
{CHIME/FRB Collaboration}, {Andersen}, B.~C., {Bandura}, K.~M., {Bhardwaj}, M.,
  {Bij}, A., {Boyce}, M.~M., et~al. (2020).
\newblock {A bright millisecond-duration radio burst from a Galactic magnetar}.
\newblock \emph{\nat} 587, 54--58.
\newblock \doi{10.1038/s41586-020-2863-y}
\bibAnnoteFile{chime20}

\bibitem[{{Cowperthwaite} et~al.(2017){Cowperthwaite}, {Berger}, {Villar},
  {Metzger}, {Nicholl}, {Chornock} et~al.}]{cowperthwaite17}
{Cowperthwaite}, P.~S., {Berger}, E., {Villar}, V.~A., {Metzger}, B.~D.,
  {Nicholl}, M., {Chornock}, R., et~al. (2017).
\newblock {The Electromagnetic Counterpart of the Binary Neutron Star Merger
  LIGO/Virgo GW170817. II. UV, Optical, and Near-infrared Light Curves and
  Comparison to Kilonova Models}.
\newblock \emph{\apjl} 848, L17.
\newblock \doi{10.3847/2041-8213/aa8fc7}
\bibAnnoteFile{cowperthwaite17}

\bibitem[{{Dai} et~al.(2018){Dai}, {McKinney}, {Roth}, {Ramirez-Ruiz}, and
  {Miller}}]{dai18}
{Dai}, L., {McKinney}, J.~C., {Roth}, N., {Ramirez-Ruiz}, E., and {Miller},
  M.~C. (2018).
\newblock {A Unified Model for Tidal Disruption Events}.
\newblock \emph{\apjl} 859, L20.
\newblock \doi{10.3847/2041-8213/aab429}
\bibAnnoteFile{dai18}

\bibitem[{{D'Avanzo} et~al.(2018){D'Avanzo}, {Campana}, {Salafia}, {Ghirlanda},
  {Ghisellini}, {Melandri} et~al.}]{davanzo18}
{D'Avanzo}, P., {Campana}, S., {Salafia}, O.~S., {Ghirlanda}, G., {Ghisellini},
  G., {Melandri}, A., et~al. (2018).
\newblock {The evolution of the X-ray afterglow emission of GW 170817/ GRB
  170817A in XMM-Newton observations}.
\newblock \emph{\aap} 613, L1.
\newblock \doi{10.1051/0004-6361/201832664}
\bibAnnoteFile{davanzo18}

\bibitem[{{Dewdney} et~al.(2009){Dewdney}, {Hall}, {Schilizzi}, and
  {Lazio}}]{dewdney09}
{Dewdney}, P.~E., {Hall}, P.~J., {Schilizzi}, R.~T., and {Lazio}, T.~J.~L.~W.
  (2009).
\newblock {The Square Kilometre Array}.
\newblock \emph{IEEE Proceedings} 97, 1482--1496.
\newblock \doi{10.1109/JPROC.2009.2021005}
\bibAnnoteFile{dewdney09}

\bibitem[{{Drout} et~al.(2014){Drout}, {Chornock}, {Soderberg}, {Sanders},
  {McKinnon}, {Rest} et~al.}]{drout14}
{Drout}, M.~R., {Chornock}, R., {Soderberg}, A.~M., {Sanders}, N.~E.,
  {McKinnon}, R., {Rest}, A., et~al. (2014).
\newblock {Rapidly Evolving and Luminous Transients from Pan-STARRS1}.
\newblock \emph{\apj} 794, 23.
\newblock \doi{10.1088/0004-637X/794/1/23}
\bibAnnoteFile{drout14}

\bibitem[{{Eraerds} et~al.(2021){Eraerds}, {Antonelli}, {Davis}, {Hall},
  {Hetherington}, {Holland} et~al.}]{Eraerds2021}
{Eraerds}, T., {Antonelli}, V., {Davis}, C., {Hall}, D., {Hetherington}, O.,
  {Holland}, A., et~al. (2021).
\newblock {Enhanced simulations on the Athena/Wide Field Imager instrumental
  background}.
\newblock \emph{Journal of Astronomical Telescopes, Instruments, and Systems}
  7, 034001.
\newblock \doi{10.1117/1.JATIS.7.3.034001}
\bibAnnoteFile{Eraerds2021}

\bibitem[{{Evans} and {Kochanek}(1989)}]{evans89}
{Evans}, C.~R. and {Kochanek}, C.~S. (1989).
\newblock {The Tidal Disruption of a Star by a Massive Black Hole}.
\newblock \emph{\apjl} 346, L13.
\newblock \doi{10.1086/185567}
\bibAnnoteFile{evans89}

\bibitem[{{Gezari}(2021)}]{gezari21}
{Gezari}, S. (2021).
\newblock {Tidal Disruption Events}.
\newblock \emph{\araa} 59, 21--58.
\newblock \doi{10.1146/annurev-astro-111720-030029}
\bibAnnoteFile{gezari21}

\bibitem[{{Graham} et~al.(2020){Graham}, {Ford}, {McKernan}, {Ross}, {Stern},
  {Burdge} et~al.}]{graham20}
{Graham}, M.~J., {Ford}, K.~E.~S., {McKernan}, B., {Ross}, N.~P., {Stern}, D.,
  {Burdge}, K., et~al. (2020).
\newblock {Candidate Electromagnetic Counterpart to the Binary Black Hole
  Merger Gravitational-Wave Event S190521g$^{*}$}.
\newblock \emph{\prl} 124, 251102.
\newblock \doi{10.1103/PhysRevLett.124.251102}
\bibAnnoteFile{graham20}

\bibitem[{{Grefenstette} et~al.(2023){Grefenstette}, {Brightman}, {Earnshaw},
  {Harrison}, and {Margutti}}]{grefenstette23}
{Grefenstette}, B.~W., {Brightman}, M., {Earnshaw}, H.~P., {Harrison}, F.~A.,
  and {Margutti}, R. (2023).
\newblock {Early Hard X-Rays from the Nearby Core-collapse Supernova SN
  2023ixf}.
\newblock \emph{\apjl} 952, L3.
\newblock \doi{10.3847/2041-8213/acdf4e}
\bibAnnoteFile{grefenstette23}

\bibitem[{{Guolo} et~al.(2023){Guolo}, {Gezari}, {Yao}, {van Velzen},
  {Hammerstein}, {Cenko} et~al.}]{guolo23}
{Guolo}, M., {Gezari}, S., {Yao}, Y., {van Velzen}, S., {Hammerstein}, E.,
  {Cenko}, S.~B., et~al. (2023).
\newblock {A systematic analysis of the X-ray emission in optically selected
  tidal disruption events: observational evidence for the unification of the
  optically and X-ray selected populations}.
\newblock \emph{arXiv e-prints} ,
  arXiv:2308.13019\doi{10.48550/arXiv.2308.13019}
\bibAnnoteFile{guolo23}

\bibitem[{{Hajela} et~al.(2022){Hajela}, {Margutti}, {Bright}, {Alexander},
  {Metzger}, {Nedora} et~al.}]{hajela22}
{Hajela}, A., {Margutti}, R., {Bright}, J.~S., {Alexander}, K.~D., {Metzger},
  B.~D., {Nedora}, V., et~al. (2022).
\newblock {Evidence for X-Ray Emission in Excess to the Jet-afterglow Decay 3.5
  yr after the Binary Neutron Star Merger GW 170817: A New Emission Component}.
\newblock \emph{\apjl} 927, L17.
\newblock \doi{10.3847/2041-8213/ac504a}
\bibAnnoteFile{hajela22}

\bibitem[{{Harrison} et~al.(2013){Harrison}, {Craig}, {Christensen}, {Hailey},
  {Zhang}, {Boggs} et~al.}]{Harrison2013}
{Harrison}, F.~A., {Craig}, W.~W., {Christensen}, F.~E., {Hailey}, C.~J.,
  {Zhang}, W.~W., {Boggs}, S.~E., et~al. (2013).
\newblock {The Nuclear Spectroscopic Telescope Array (NuSTAR) High-energy X-Ray
  Mission}.
\newblock \emph{\apj} 770, 103.
\newblock \doi{10.1088/0004-637X/770/2/103}
\bibAnnoteFile{Harrison2013}

\bibitem[{{Ho} et~al.(2021){Ho}, {Perley}, {Gal-Yam}, {Lunnan}, {Sollerman},
  {Schulze} et~al.}]{ho21}
{Ho}, A. Y.~Q., {Perley}, D.~A., {Gal-Yam}, A., {Lunnan}, R., {Sollerman}, J.,
  {Schulze}, S., et~al. (2021).
\newblock {A Search for Extragalactic Fast Blue Optical Transients in ZTF and
  the Rate of AT2018cow-like Transients}.
\newblock \emph{arXiv e-prints} ,
  arXiv:2105.08811\doi{10.48550/arXiv.2105.08811}
\bibAnnoteFile{ho21}

\bibitem[{{Ho} et~al.(2023){Ho}, {Perley}, {Gal-Yam}, {Lunnan}, {Sollerman},
  {Schulze} et~al.}]{ho23}
{Ho}, A. Y.~Q., {Perley}, D.~A., {Gal-Yam}, A., {Lunnan}, R., {Sollerman}, J.,
  {Schulze}, S., et~al. (2023).
\newblock {A Search for Extragalactic Fast Blue Optical Transients in ZTF and
  the Rate of AT2018cow-like Transients}.
\newblock \emph{\apj} 949, 120.
\newblock \doi{10.3847/1538-4357/acc533}
\bibAnnoteFile{ho23}

\bibitem[{{Hovatta} et~al.(2009){Hovatta}, {Valtaoja}, {Tornikoski}, and
  {L{\"a}hteenm{\"a}ki}}]{hovatta09}
{Hovatta}, T., {Valtaoja}, E., {Tornikoski}, M., and {L{\"a}hteenm{\"a}ki}, A.
  (2009).
\newblock {Doppler factors, Lorentz factors and viewing angles for quasars, BL
  Lacertae objects and radio galaxies}.
\newblock \emph{\aap} 494, 527--537.
\newblock \doi{10.1051/0004-6361:200811150}
\bibAnnoteFile{hovatta09}

\bibitem[{{IceCube Collaboration} et~al.(2018){IceCube Collaboration},
  {Aartsen}, {Ackermann}, {Adams}, {Aguilar}, {Ahlers} et~al.}]{icecube18}
{IceCube Collaboration}, {Aartsen}, M.~G., {Ackermann}, M., {Adams}, J.,
  {Aguilar}, J.~A., {Ahlers}, M., et~al. (2018).
\newblock {Multimessenger observations of a flaring blazar coincident with
  high-energy neutrino IceCube-170922A}.
\newblock \emph{Science} 361, eaat1378.
\newblock \doi{10.1126/science.aat1378}
\bibAnnoteFile{icecube18}

\bibitem[{{IceCube-Gen2 Collaboration} et~al.(2014){IceCube-Gen2
  Collaboration}, {:}, {Aartsen}, {Ackermann}, {Adams}, {Aguilar}
  et~al.}]{icecube2}
{IceCube-Gen2 Collaboration}, {:}, {Aartsen}, M.~G., {Ackermann}, M., {Adams},
  J., {Aguilar}, J.~A., et~al. (2014).
\newblock {IceCube-Gen2: A Vision for the Future of Neutrino Astronomy in
  Antarctica}.
\newblock \emph{arXiv e-prints} , arXiv:1412.5106\doi{10.48550/arXiv.1412.5106}
\bibAnnoteFile{icecube2}

\bibitem[{{Ivezi{\'c}} et~al.(2019){Ivezi{\'c}}, {Kahn}, {Tyson}, {Abel},
  {Acosta}, {Allsman} et~al.}]{ivezic19}
{Ivezi{\'c}}, {\v{Z}}., {Kahn}, S.~M., {Tyson}, J.~A., {Abel}, B., {Acosta},
  E., {Allsman}, R., et~al. (2019).
\newblock {LSST: From Science Drivers to Reference Design and Anticipated Data
  Products}.
\newblock \emph{\apj} 873, 111.
\newblock \doi{10.3847/1538-4357/ab042c}
\bibAnnoteFile{ivezic19}

\bibitem[{{Jansen} et~al.(2001){Jansen}, {Lumb}, {Altieri}, {Clavel}, {Ehle},
  {Erd} et~al.}]{Jansen2001}
{Jansen}, F., {Lumb}, D., {Altieri}, B., {Clavel}, J., {Ehle}, M., {Erd}, C.,
  et~al. (2001).
\newblock {XMM-Newton observatory. I. The spacecraft and operations}.
\newblock \emph{\aap} 365, L1--L6.
\newblock \doi{10.1051/0004-6361:20000036}
\bibAnnoteFile{Jansen2001}

\bibitem[{{Johnston} et~al.(2008){Johnston}, {Taylor}, {Bailes}, {Bartel},
  {Baugh}, {Bietenholz} et~al.}]{johnston08}
{Johnston}, S., {Taylor}, R., {Bailes}, M., {Bartel}, N., {Baugh}, C.,
  {Bietenholz}, M., et~al. (2008).
\newblock {Science with ASKAP. The Australian square-kilometre-array
  pathfinder}.
\newblock \emph{Experimental Astronomy} 22, 151--273.
\newblock \doi{10.1007/s10686-008-9124-7}
\bibAnnoteFile{johnston08}

\bibitem[{{Kara} et~al.(2016){Kara}, {Miller}, {Reynolds}, and {Dai}}]{kara16}
{Kara}, E., {Miller}, J.~M., {Reynolds}, C., and {Dai}, L. (2016).
\newblock {Relativistic reverberation in the accretion flow of a tidal
  disruption event}.
\newblock \emph{\nat} 535, 388--390.
\newblock \doi{10.1038/nature18007}
\bibAnnoteFile{kara16}

\bibitem[{{Katz}(2006)}]{katz06}
{Katz}, U.~F. (2006).
\newblock {KM3NeT: Towards a km$^{3}$ Mediterranean neutrino telescope}.
\newblock \emph{Nuclear Instruments and Methods in Physics Research A} 567,
  457--461.
\newblock \doi{10.1016/j.nima.2006.05.235}
\bibAnnoteFile{katz06}

\bibitem[{{Kimura} et~al.(2021){Kimura}, {Murase}, and {Bartos}}]{kimura21}
{Kimura}, S.~S., {Murase}, K., and {Bartos}, I. (2021).
\newblock {Outflow Bubbles from Compact Binary Mergers Embedded in Active
  Galactic Nuclei: Cavity Formation and the Impact on Electromagnetic
  Counterparts}.
\newblock \emph{\apj} 916, 111.
\newblock \doi{10.3847/1538-4357/ac0535}
\bibAnnoteFile{kimura21}

\bibitem[{{Laha} et~al.(2022){Laha}, {Meyer}, {Roychowdhury}, {Becerra
  Gonzalez}, {Acosta-Pulido}, {Thapa} et~al.}]{laha22}
{Laha}, S., {Meyer}, E., {Roychowdhury}, A., {Becerra Gonzalez}, J.,
  {Acosta-Pulido}, J.~A., {Thapa}, A., et~al. (2022).
\newblock {A Radio, Optical, UV, and X-Ray View of the Enigmatic Changing-look
  Active Galactic Nucleus 1ES 1927+654 from Its Pre- to Postflare States}.
\newblock \emph{\apj} 931, 5.
\newblock \doi{10.3847/1538-4357/ac63aa}
\bibAnnoteFile{laha22}

\bibitem[{{Lazzati} et~al.(2017){Lazzati}, {L{\'o}pez-C{\'a}mara}, {Cantiello},
  {Morsony}, {Perna}, and {Workman}}]{lazzati17}
{Lazzati}, D., {L{\'o}pez-C{\'a}mara}, D., {Cantiello}, M., {Morsony}, B.~J.,
  {Perna}, R., and {Workman}, J.~C. (2017).
\newblock {Off-axis Prompt X-Ray Transients from the Cocoon of Short Gamma-Ray
  Bursts}.
\newblock \emph{\apjl} 848, L6.
\newblock \doi{10.3847/2041-8213/aa8f3d}
\bibAnnoteFile{lazzati17}

\bibitem[{{Lazzati} et~al.(2018){Lazzati}, {Perna}, {Morsony}, {Lopez-Camara},
  {Cantiello}, {Ciolfi} et~al.}]{lazzati18}
{Lazzati}, D., {Perna}, R., {Morsony}, B.~J., {Lopez-Camara}, D., {Cantiello},
  M., {Ciolfi}, R., et~al. (2018).
\newblock {Late Time Afterglow Observations Reveal a Collimated Relativistic
  Jet in the Ejecta of the Binary Neutron Star Merger GW170817}.
\newblock \emph{\prl} 120, 241103.
\newblock \doi{10.1103/PhysRevLett.120.241103}
\bibAnnoteFile{lazzati18}

\bibitem[{{Li} et~al.(2021){Li}, {Lin}, {Xiong}, {Ge}, {Li}, {Li}
  et~al.}]{li21}
{Li}, C.~K., {Lin}, L., {Xiong}, S.~L., {Ge}, M.~Y., {Li}, X.~B., {Li}, T.~P.,
  et~al. (2021).
\newblock {HXMT identification of a non-thermal X-ray burst from SGR J1935+2154
  and with FRB 200428}.
\newblock \emph{Nature Astronomy} 5, 378--384.
\newblock \doi{10.1038/s41550-021-01302-6}
\bibAnnoteFile{li21}

\bibitem[{{Liu} et~al.(2023){Liu}, {Malyali}, {Krumpe}, {Homan}, {Goodwin},
  {Grotova} et~al.}]{liu23}
{Liu}, Z., {Malyali}, A., {Krumpe}, M., {Homan}, D., {Goodwin}, A.~J.,
  {Grotova}, I., et~al. (2023).
\newblock {Deciphering the extreme X-ray variability of the nuclear transient
  eRASSt J045650.3{\ensuremath{-}}203750. A likely repeating partial tidal
  disruption event}.
\newblock \emph{\aap} 669, A75.
\newblock \doi{10.1051/0004-6361/202244805}
\bibAnnoteFile{liu23}

\bibitem[{{Maggiore} et~al.(2020){Maggiore}, {Van Den Broeck}, {Bartolo},
  {Belgacem}, {Bertacca}, {Bizouard} et~al.}]{maggiore20}
{Maggiore}, M., {Van Den Broeck}, C., {Bartolo}, N., {Belgacem}, E.,
  {Bertacca}, D., {Bizouard}, M.~A., et~al. (2020).
\newblock {Science case for the Einstein telescope}.
\newblock \emph{\jcap} 2020, 050.
\newblock \doi{10.1088/1475-7516/2020/03/050}
\bibAnnoteFile{maggiore20}

\bibitem[{{Mapelli}(2021)}]{mapelli21}
{Mapelli}, M. (2021).
\newblock {Formation Channels of Single and Binary Stellar-Mass Black Holes}.
\newblock In \emph{Handbook of Gravitational Wave Astronomy}. 16.
\newblock \doi{10.1007/978-981-15-4702-7_16-1}
\bibAnnoteFile{mapelli21}

\bibitem[{{Margutti} et~al.(2018){Margutti}, {Alexander}, {Xie}, {Sironi},
  {Metzger}, {Kathirgamaraju} et~al.}]{margutti18}
{Margutti}, R., {Alexander}, K.~D., {Xie}, X., {Sironi}, L., {Metzger}, B.~D.,
  {Kathirgamaraju}, A., et~al. (2018).
\newblock {The Binary Neutron Star Event LIGO/Virgo GW170817 160 Days after
  Merger: Synchrotron Emission across the Electromagnetic Spectrum}.
\newblock \emph{\apjl} 856, L18.
\newblock \doi{10.3847/2041-8213/aab2ad}
\bibAnnoteFile{margutti18}

\bibitem[{{Margutti} et~al.(2017{\natexlab{a}}){Margutti}, {Berger}, {Fong},
  {Guidorzi}, {Alexander}, {Metzger} et~al.}]{margutti17}
{Margutti}, R., {Berger}, E., {Fong}, W., {Guidorzi}, C., {Alexander}, K.~D.,
  {Metzger}, B.~D., et~al. (2017{\natexlab{a}}).
\newblock {The Electromagnetic Counterpart of the Binary Neutron Star Merger
  LIGO/Virgo GW170817. V. Rising X-Ray Emission from an Off-axis Jet}.
\newblock \emph{\apjl} 848, L20.
\newblock \doi{10.3847/2041-8213/aa9057}
\bibAnnoteFile{margutti17}

\bibitem[{{Margutti} and {Chornock}(2021)}]{margutti21}
{Margutti}, R. and {Chornock}, R. (2021).
\newblock {First Multimessenger Observations of a Neutron Star Merger}.
\newblock \emph{\araa} 59, 155--202.
\newblock \doi{10.1146/annurev-astro-112420-030742}
\bibAnnoteFile{margutti21}

\bibitem[{{Margutti} et~al.(2017{\natexlab{b}}){Margutti}, {Kamble},
  {Milisavljevic}, {Zapartas}, {de Mink}, {Drout} et~al.}]{margutti17a}
{Margutti}, R., {Kamble}, A., {Milisavljevic}, D., {Zapartas}, E., {de Mink},
  S.~E., {Drout}, M., et~al. (2017{\natexlab{b}}).
\newblock {Ejection of the Massive Hydrogen-rich Envelope Timed with the
  Collapse of the Stripped SN 2014C}.
\newblock \emph{\apj} 835, 140.
\newblock \doi{10.3847/1538-4357/835/2/140}
\bibAnnoteFile{margutti17a}

\bibitem[{{Margutti} et~al.(2019){Margutti}, {Metzger}, {Chornock}, {Vurm},
  {Roth}, {Grefenstette} et~al.}]{margutti19}
{Margutti}, R., {Metzger}, B.~D., {Chornock}, R., {Vurm}, I., {Roth}, N.,
  {Grefenstette}, B.~W., et~al. (2019).
\newblock {An Embedded X-Ray Source Shines through the Aspherical AT 2018cow:
  Revealing the Inner Workings of the Most Luminous Fast-evolving Optical
  Transients}.
\newblock \emph{\apj} 872, 18.
\newblock \doi{10.3847/1538-4357/aafa01}
\bibAnnoteFile{margutti19}

\bibitem[{{Masterson} et~al.(2022){Masterson}, {Kara}, {Ricci}, {Garc{\'\i}a},
  {Fabian}, {Pinto} et~al.}]{masterson22}
{Masterson}, M., {Kara}, E., {Ricci}, C., {Garc{\'\i}a}, J.~A., {Fabian},
  A.~C., {Pinto}, C., et~al. (2022).
\newblock {Evolution of a Relativistic Outflow and X-Ray Corona in the Extreme
  Changing-look AGN 1ES 1927+654}.
\newblock \emph{\apj} 934, 35.
\newblock \doi{10.3847/1538-4357/ac76c0}
\bibAnnoteFile{masterson22}

\bibitem[{{McKernan} et~al.(2019){McKernan}, {Ford}, {Bartos}, {Graham},
  {Lyra}, {Marka} et~al.}]{mckernan19}
{McKernan}, B., {Ford}, K.~E.~S., {Bartos}, I., {Graham}, M.~J., {Lyra}, W.,
  {Marka}, S., et~al. (2019).
\newblock {Ram-pressure Stripping of a Kicked Hill Sphere: Prompt
  Electromagnetic Emission from the Merger of Stellar Mass Black Holes in an
  AGN Accretion Disk}.
\newblock \emph{\apjl} 884, L50.
\newblock \doi{10.3847/2041-8213/ab4886}
\bibAnnoteFile{mckernan19}

\bibitem[{{McKernan} et~al.(2022){McKernan}, {Ford}, {Callister}, {Farr},
  {O'Shaughnessy}, {Smith} et~al.}]{mckernan22}
{McKernan}, B., {Ford}, K.~E.~S., {Callister}, T., {Farr}, W.~M.,
  {O'Shaughnessy}, R., {Smith}, R., et~al. (2022).
\newblock {LIGO-Virgo correlations between mass ratio and effective inspiral
  spin: testing the active galactic nuclei channel}.
\newblock \emph{\mnras} 514, 3886--3893.
\newblock \doi{10.1093/mnras/stac1570}
\bibAnnoteFile{mckernan22}

\bibitem[{{Meidinger} et~al.(2020){Meidinger}, {Albrecht}, {Beitler},
  {Bonholzer}, {Emberger}, {Frank} et~al.}]{meidinger20}
{Meidinger}, N., {Albrecht}, S., {Beitler}, C., {Bonholzer}, M., {Emberger},
  V., {Frank}, J., et~al. (2020).
\newblock {Development status of the wide field imager instrument for Athena}.
\newblock In \emph{Society of Photo-Optical Instrumentation Engineers (SPIE)
  Conference Series}. vol. 11444 of \emph{Society of Photo-Optical
  Instrumentation Engineers (SPIE) Conference Series}, 114440T.
\newblock \doi{10.1117/12.2560507}
\bibAnnoteFile{meidinger20}

\bibitem[{{Mereghetti} et~al.(2020){Mereghetti}, {Savchenko}, {Ferrigno},
  {G{\"o}tz}, {Rigoselli}, {Tiengo} et~al.}]{mereghetti20}
{Mereghetti}, S., {Savchenko}, V., {Ferrigno}, C., {G{\"o}tz}, D., {Rigoselli},
  M., {Tiengo}, A., et~al. (2020).
\newblock {INTEGRAL Discovery of a Burst with Associated Radio Emission from
  the Magnetar SGR 1935+2154}.
\newblock \emph{\apjl} 898, L29.
\newblock \doi{10.3847/2041-8213/aba2cf}
\bibAnnoteFile{mereghetti20}

\bibitem[{{Metzger} and {Fern{\'a}ndez}(2014)}]{metzger14}
{Metzger}, B.~D. and {Fern{\'a}ndez}, R. (2014).
\newblock {Red or blue? A potential kilonova imprint of the delay until black
  hole formation following a neutron star merger}.
\newblock \emph{\mnras} 441, 3444--3453.
\newblock \doi{10.1093/mnras/stu802}
\bibAnnoteFile{metzger14}

\bibitem[{{Mooley} et~al.(2022){Mooley}, {Anderson}, and {Lu}}]{mooley22}
{Mooley}, K.~P., {Anderson}, J., and {Lu}, W. (2022).
\newblock {Optical superluminal motion measurement in the neutron-star merger
  GW170817}.
\newblock \emph{\nat} 610, 273--276.
\newblock \doi{10.1038/s41586-022-05145-7}
\bibAnnoteFile{mooley22}

\bibitem[{{Mooley} et~al.(2018{\natexlab{a}}){Mooley}, {Deller}, {Gottlieb},
  {Nakar}, {Hallinan}, {Bourke} et~al.}]{mooley18b}
{Mooley}, K.~P., {Deller}, A.~T., {Gottlieb}, O., {Nakar}, E., {Hallinan}, G.,
  {Bourke}, S., et~al. (2018{\natexlab{a}}).
\newblock {Superluminal motion of a relativistic jet in the neutron-star merger
  GW170817}.
\newblock \emph{\nat} 561, 355--359.
\newblock \doi{10.1038/s41586-018-0486-3}
\bibAnnoteFile{mooley18b}

\bibitem[{{Mooley} et~al.(2018{\natexlab{b}}){Mooley}, {Nakar}, {Hotokezaka},
  {Hallinan}, {Corsi}, {Frail} et~al.}]{mooley18}
{Mooley}, K.~P., {Nakar}, E., {Hotokezaka}, K., {Hallinan}, G., {Corsi}, A.,
  {Frail}, D.~A., et~al. (2018{\natexlab{b}}).
\newblock {A mildly relativistic wide-angle outflow in the neutron-star merger
  event GW170817}.
\newblock \emph{\nat} 554, 207--210.
\newblock \doi{10.1038/nature25452}
\bibAnnoteFile{mooley18}

\bibitem[{{Murase} et~al.(2018{\natexlab{a}}){Murase}, {Oikonomou}, and
  {Petropoulou}}]{murase18a}
{Murase}, K., {Oikonomou}, F., and {Petropoulou}, M. (2018{\natexlab{a}}).
\newblock {Blazar Flares as an Origin of High-energy Cosmic Neutrinos?}
\newblock \emph{\apj} 865, 124.
\newblock \doi{10.3847/1538-4357/aada00}
\bibAnnoteFile{murase18a}

\bibitem[{{Murase} et~al.(2018{\natexlab{b}}){Murase}, {Toomey}, {Fang},
  {Oikonomou}, {Kimura}, {Hotokezaka} et~al.}]{murase18}
{Murase}, K., {Toomey}, M.~W., {Fang}, K., {Oikonomou}, F., {Kimura}, S.~S.,
  {Hotokezaka}, K., et~al. (2018{\natexlab{b}}).
\newblock {Double Neutron Star Mergers and Short Gamma-ray Bursts: Long-lasting
  High-energy Signatures and Remnant Dichotomy}.
\newblock \emph{\apj} 854, 60.
\newblock \doi{10.3847/1538-4357/aaa48a}
\bibAnnoteFile{murase18}

\bibitem[{{Nandra} et~al.(2013){Nandra}, {Barret}, {Barcons}, {Fabian}, {den
  Herder}, {Piro} et~al.}]{nandra13}
{Nandra}, K., {Barret}, D., {Barcons}, X., {Fabian}, A., {den Herder}, J.-W.,
  {Piro}, L., et~al. (2013).
\newblock {The Hot and Energetic Universe: A White Paper presenting the science
  theme motivating the Athena+ mission}.
\newblock \emph{ArXiv e-prints}
\bibAnnoteFile{nandra13}

\bibitem[{{Nathanail} et~al.(2021){Nathanail}, {Gill}, {Porth}, {Fromm}, and
  {Rezzolla}}]{nathanail21}
{Nathanail}, A., {Gill}, R., {Porth}, O., {Fromm}, C.~M., and {Rezzolla}, L.
  (2021).
\newblock {3D magnetized jet break-out from neutron-star binary merger ejecta:
  afterglow emission from the jet and the ejecta}.
\newblock \emph{\mnras} 502, 1843--1855.
\newblock \doi{10.1093/mnras/stab115}
\bibAnnoteFile{nathanail21}

\bibitem[{{Ofek} et~al.(2014){Ofek}, {Zoglauer}, {Boggs}, {Barri{\'e}re},
  {Reynolds}, {Fryer} et~al.}]{ofek14}
{Ofek}, E.~O., {Zoglauer}, A., {Boggs}, S.~E., {Barri{\'e}re}, N.~M.,
  {Reynolds}, S.~P., {Fryer}, C.~L., et~al. (2014).
\newblock {SN 2010jl: Optical to Hard X-Ray Observations Reveal an Explosion
  Embedded in a Ten Solar Mass Cocoon}.
\newblock \emph{\apj} 781, 42.
\newblock \doi{10.1088/0004-637X/781/1/42}
\bibAnnoteFile{ofek14}

\bibitem[{{Pasham} et~al.(2023){Pasham}, {Lucchini}, {Laskar}, {Gompertz},
  {Srivastav}, {Nicholl} et~al.}]{pasham23}
{Pasham}, D.~R., {Lucchini}, M., {Laskar}, T., {Gompertz}, B.~P., {Srivastav},
  S., {Nicholl}, M., et~al. (2023).
\newblock {The Birth of a Relativistic Jet Following the Disruption of a Star
  by a Cosmological Black Hole}.
\newblock \emph{Nature Astronomy} 7, 88--104.
\newblock \doi{10.1038/s41550-022-01820-x}
\bibAnnoteFile{pasham23}

\bibitem[{{Petroff} et~al.(2019){Petroff}, {Hessels}, and
  {Lorimer}}]{petroff19}
{Petroff}, E., {Hessels}, J.~W.~T., and {Lorimer}, D.~R. (2019).
\newblock {Fast radio bursts}.
\newblock \emph{\aapr} 27, 4.
\newblock \doi{10.1007/s00159-019-0116-6}
\bibAnnoteFile{petroff19}

\bibitem[{{Polzin} et~al.(2022){Polzin}, {Margutti}, {Coppejans}, {Auchettl},
  {Page}, {Vasilopoulos} et~al.}]{polzin22}
{Polzin}, A., {Margutti}, R., {Coppejans}, D., {Auchettl}, K., {Page}, K.~L.,
  {Vasilopoulos}, G., et~al. (2022).
\newblock {The Luminosity Phase Space of Galactic and Extragalactic X-ray
  Transients Out to Intermediate Redshifts}.
\newblock \emph{arXiv e-prints} , arXiv:2211.01232
\bibAnnoteFile{polzin22}

\bibitem[{{Punturo} et~al.(2010){Punturo}, {Abernathy}, {Acernese}, {Allen},
  {Andersson}, {Arun} et~al.}]{punturo10}
{Punturo}, M., {Abernathy}, M., {Acernese}, F., {Allen}, B., {Andersson}, N.,
  {Arun}, K., et~al. (2010).
\newblock {The Einstein Telescope: a third-generation gravitational wave
  observatory}.
\newblock \emph{Classical and Quantum Gravity} 27, 194002.
\newblock \doi{10.1088/0264-9381/27/19/194002}
\bibAnnoteFile{punturo10}

\bibitem[{{Rees}(1988)}]{rees88}
{Rees}, M.~J. (1988).
\newblock {Tidal disruption of stars by black holes of 10 to the 6th-10 to the
  8th solar masses in nearby galaxies}.
\newblock \emph{\nat} 333, 523--528.
\newblock \doi{10.1038/333523a0}
\bibAnnoteFile{rees88}

\bibitem[{{Reimer} et~al.(2019){Reimer}, {B{\"o}ttcher}, and
  {Buson}}]{reimer19}
{Reimer}, A., {B{\"o}ttcher}, M., and {Buson}, S. (2019).
\newblock {Cascading Constraints from Neutrino-emitting Blazars: The Case of
  TXS 0506+056}.
\newblock \emph{\apj} 881, 46.
\newblock \doi{10.3847/1538-4357/ab2bff}
\bibAnnoteFile{reimer19}

\bibitem[{{Reitze} et~al.(2019){Reitze}, {Adhikari}, {Ballmer}, {Barish},
  {Barsotti}, {Billingsley} et~al.}]{reitze19}
{Reitze}, D., {Adhikari}, R.~X., {Ballmer}, S., {Barish}, B., {Barsotti}, L.,
  {Billingsley}, G., et~al. (2019).
\newblock {Cosmic Explorer: The U.S. Contribution to Gravitational-Wave
  Astronomy beyond LIGO}.
\newblock In \emph{Bulletin of the American Astronomical Society}. vol.~51, 35.
\newblock \doi{10.48550/arXiv.1907.04833}
\bibAnnoteFile{reitze19}

\bibitem[{{Remillard} and {McClintock}(2006)}]{remillard06}
{Remillard}, R.~A. and {McClintock}, J.~E. (2006).
\newblock {X-Ray Properties of Black-Hole Binaries}.
\newblock \emph{\araa} 44, 49--92.
\newblock \doi{10.1146/annurev.astro.44.051905.092532}
\bibAnnoteFile{remillard06}

\bibitem[{{Reusch} et~al.(2022){Reusch}, {Stein}, {Kowalski}, {van Velzen},
  {Franckowiak}, {Lunardini} et~al.}]{reusch22}
{Reusch}, S., {Stein}, R., {Kowalski}, M., {van Velzen}, S., {Franckowiak}, A.,
  {Lunardini}, C., et~al. (2022).
\newblock {Candidate Tidal Disruption Event AT2019fdr Coincident with a
  High-Energy Neutrino}.
\newblock \emph{\prl} 128, 221101.
\newblock \doi{10.1103/PhysRevLett.128.221101}
\bibAnnoteFile{reusch22}

\bibitem[{{Ricci} et~al.(2020){Ricci}, {Kara}, {Loewenstein}, {Trakhtenbrot},
  {Arcavi}, {Remillard} et~al.}]{ricci20}
{Ricci}, C., {Kara}, E., {Loewenstein}, M., {Trakhtenbrot}, B., {Arcavi}, I.,
  {Remillard}, R., et~al. (2020).
\newblock {The Destruction and Recreation of the X-Ray Corona in a
  Changing-look Active Galactic Nucleus}.
\newblock \emph{\apjl} 898, L1.
\newblock \doi{10.3847/2041-8213/ab91a1}
\bibAnnoteFile{ricci20}

\bibitem[{{Ricci} et~al.(2021){Ricci}, {Loewenstein}, {Kara}, {Remillard},
  {Trakhtenbrot}, {Arcavi} et~al.}]{ricci21}
{Ricci}, C., {Loewenstein}, M., {Kara}, E., {Remillard}, R., {Trakhtenbrot},
  B., {Arcavi}, I., et~al. (2021).
\newblock {The 450 Day X-Ray Monitoring of the Changing-look AGN 1ES 1927+654}.
\newblock \emph{\apjs} 255, 7.
\newblock \doi{10.3847/1538-4365/abe94b}
\bibAnnoteFile{ricci21}

\bibitem[{{Ricci} and {Trakhtenbrot}(2022)}]{ricci22}
{Ricci}, C. and {Trakhtenbrot}, B. (2022).
\newblock {Changing-look Active Galactic Nuclei}.
\newblock \emph{arXiv e-prints} ,
  arXiv:2211.05132\doi{10.48550/arXiv.2211.05132}
\bibAnnoteFile{ricci22}

\bibitem[{{Ridnaia} et~al.(2021){Ridnaia}, {Svinkin}, {Frederiks}, {Bykov},
  {Popov}, {Aptekar} et~al.}]{ridnaia21}
{Ridnaia}, A., {Svinkin}, D., {Frederiks}, D., {Bykov}, A., {Popov}, S.,
  {Aptekar}, R., et~al. (2021).
\newblock {A peculiar hard X-ray counterpart of a Galactic fast radio burst}.
\newblock \emph{Nature Astronomy} 5, 372--377.
\newblock \doi{10.1038/s41550-020-01265-0}
\bibAnnoteFile{ridnaia21}

\bibitem[{{Scepi} et~al.(2021){Scepi}, {Begelman}, and {Dexter}}]{scepi21}
{Scepi}, N., {Begelman}, M.~C., and {Dexter}, J. (2021).
\newblock {Magnetic flux inversion in a peculiar changing look AGN}.
\newblock \emph{\mnras} 502, L50--L54.
\newblock \doi{10.1093/mnrasl/slab002}
\bibAnnoteFile{scepi21}

\bibitem[{{Shvartzvald} et~al.(2023){Shvartzvald}, {Waxman}, {Gal-Yam}, {Ofek},
  {Ben-Ami}, {Berge} et~al.}]{shvartzvald23}
{Shvartzvald}, Y., {Waxman}, E., {Gal-Yam}, A., {Ofek}, E.~O., {Ben-Ami}, S.,
  {Berge}, D., et~al. (2023).
\newblock {ULTRASAT: A wide-field time-domain UV space telescope}.
\newblock \emph{arXiv e-prints} ,
  arXiv:2304.14482\doi{10.48550/arXiv.2304.14482}
\bibAnnoteFile{shvartzvald23}

\bibitem[{{Stein} et~al.(2021){Stein}, {van Velzen}, {Kowalski}, {Franckowiak},
  {Gezari}, {Miller-Jones} et~al.}]{stein21}
{Stein}, R., {van Velzen}, S., {Kowalski}, M., {Franckowiak}, A., {Gezari}, S.,
  {Miller-Jones}, J. C.~A., et~al. (2021).
\newblock {A tidal disruption event coincident with a high-energy neutrino}.
\newblock \emph{Nature Astronomy} 5, 510--518.
\newblock \doi{10.1038/s41550-020-01295-8}
\bibAnnoteFile{stein21}

\bibitem[{{Sun} et~al.(2022){Sun}, {Ruiz}, {Shapiro}, and {Tsokaros}}]{sun22}
{Sun}, L., {Ruiz}, M., {Shapiro}, S.~L., and {Tsokaros}, A. (2022).
\newblock {Jet launching from binary neutron star mergers: Incorporating
  neutrino transport and magnetic fields}.
\newblock \emph{\prd} 105, 104028.
\newblock \doi{10.1103/PhysRevD.105.104028}
\bibAnnoteFile{sun22}

\bibitem[{{Trakhtenbrot} et~al.(2019){Trakhtenbrot}, {Arcavi}, {MacLeod},
  {Ricci}, {Kara}, {Graham} et~al.}]{trakhtenbrot19}
{Trakhtenbrot}, B., {Arcavi}, I., {MacLeod}, C.~L., {Ricci}, C., {Kara}, E.,
  {Graham}, M.~L., et~al. (2019).
\newblock {1ES 1927+654: An AGN Caught Changing Look on a Timescale of Months}.
\newblock \emph{\apj} 883, 94.
\newblock \doi{10.3847/1538-4357/ab39e4}
\bibAnnoteFile{trakhtenbrot19}

\bibitem[{{Troja} et~al.(2017){Troja}, {Piro}, {van Eerten}, {Wollaeger}, {Im},
  {Fox} et~al.}]{troja17}
{Troja}, E., {Piro}, L., {van Eerten}, H., {Wollaeger}, R.~T., {Im}, M., {Fox},
  O.~D., et~al. (2017).
\newblock {The X-ray counterpart to the gravitational-wave event GW170817}.
\newblock \emph{\nat} 551, 71--74.
\newblock \doi{10.1038/nature24290}
\bibAnnoteFile{troja17}

\bibitem[{{Urry} and {Padovani}(1995)}]{urry95}
{Urry}, C.~M. and {Padovani}, P. (1995).
\newblock {Unified Schemes for Radio-Loud Active Galactic Nuclei}.
\newblock \emph{\pasp} 107, 803--+.
\newblock \doi{10.1086/133630}
\bibAnnoteFile{urry95}

\bibitem[{{Vajpeyi} et~al.(2022){Vajpeyi}, {Thrane}, {Smith}, {McKernan}, and
  {Saavik Ford}}]{vajpeyi22}
{Vajpeyi}, A., {Thrane}, E., {Smith}, R., {McKernan}, B., and {Saavik Ford},
  K.~E. (2022).
\newblock {Measuring the Properties of Active Galactic Nuclei Disks with
  Gravitational Waves}.
\newblock \emph{\apj} 931, 82.
\newblock \doi{10.3847/1538-4357/ac6180}
\bibAnnoteFile{vajpeyi22}

\bibitem[{{van Velzen} et~al.(2021){van Velzen}, {Stein}, {Gilfanov},
  {Kowalski}, {Hayasaki}, {Reusch} et~al.}]{vanvelzen21}
{van Velzen}, S., {Stein}, R., {Gilfanov}, M., {Kowalski}, M., {Hayasaki}, K.,
  {Reusch}, S., et~al. (2021).
\newblock {Establishing accretion flares from massive black holes as a major
  source of high-energy neutrinos}.
\newblock \emph{arXiv e-prints} ,
  arXiv:2111.09391\doi{10.48550/arXiv.2111.09391}
\bibAnnoteFile{vanvelzen21}

\bibitem[{{Wadiasingh} and {Timokhin}(2019)}]{wadiasingh19}
{Wadiasingh}, Z. and {Timokhin}, A. (2019).
\newblock {Repeating Fast Radio Bursts from Magnetars with Low Magnetospheric
  Twist}.
\newblock \emph{\apj} 879, 4.
\newblock \doi{10.3847/1538-4357/ab2240}
\bibAnnoteFile{wadiasingh19}

\bibitem[{{Wevers} et~al.(2021){Wevers}, {Pasham}, {van Velzen},
  {Miller-Jones}, {Uttley}, {Gendreau} et~al.}]{wevers21}
{Wevers}, T., {Pasham}, D.~R., {van Velzen}, S., {Miller-Jones}, J.~C.~A.,
  {Uttley}, P., {Gendreau}, K.~C., et~al. (2021).
\newblock {Rapid Accretion State Transitions following the Tidal Disruption
  Event AT2018fyk}.
\newblock \emph{\apj} 912, 151.
\newblock \doi{10.3847/1538-4357/abf5e2}
\bibAnnoteFile{wevers21}

\bibitem[{{Yao} et~al.(2022){Yao}, {Lu}, {Guolo}, {Pasham}, {Gezari},
  {Gilfanov} et~al.}]{yao22}
{Yao}, Y., {Lu}, W., {Guolo}, M., {Pasham}, D.~R., {Gezari}, S., {Gilfanov},
  M., et~al. (2022).
\newblock {The Tidal Disruption Event AT2021ehb: Evidence of Relativistic Disk
  Reflection, and Rapid Evolution of the Disk-Corona System}.
\newblock \emph{\apj} 937, 8.
\newblock \doi{10.3847/1538-4357/ac898a}
\bibAnnoteFile{yao22}

\bibitem[{{Younes} et~al.(2020){Younes}, {G{\"u}ver}, {Kouveliotou}, {Baring},
  {Hu}, {Wadiasingh} et~al.}]{younes20}
{Younes}, G., {G{\"u}ver}, T., {Kouveliotou}, C., {Baring}, M.~G., {Hu}, C.-P.,
  {Wadiasingh}, Z., et~al. (2020).
\newblock {NICER View of the 2020 Burst Storm and Persistent Emission of SGR
  1935+2154}.
\newblock \emph{\apjl} 904, L21.
\newblock \doi{10.3847/2041-8213/abc94c}
\bibAnnoteFile{younes20}

\bibitem[{{Zhang} et~al.(2020){Zhang}, {Petropoulou}, {Murase}, and
  {Oikonomou}}]{zhang20}
{Zhang}, B.~T., {Petropoulou}, M., {Murase}, K., and {Oikonomou}, F. (2020).
\newblock {A Neutral Beam Model for High-energy Neutrino Emission from the
  Blazar TXS 0506+056}.
\newblock \emph{\apj} 889, 118.
\newblock \doi{10.3847/1538-4357/ab659a}
\bibAnnoteFile{zhang20}

\bibitem[{{Zhao} et~al.(2023){Zhao}, {Lu}, {Chu}, and {Zhao}}]{zhao23}
{Zhao}, C., {Lu}, Y., {Chu}, Q., and {Zhao}, W. (2023).
\newblock {The luminosity functions of kilonovae from binary neutron star
  mergers under different equation of states}.
\newblock \emph{\mnras} 522, 912--936.
\newblock \doi{10.1093/mnras/stad1028}
\bibAnnoteFile{zhao23}

\end{thebibliography}

\end{document}